%% file: icfp105-johnson.tex
\documentclass[9pt]{sigplanconf} 
\usepackage{alltt,mathpartir}
\usepackage{amsmath,amsthm}
\usepackage{amssymb}
\usepackage{stmaryrd}
\usepackage{url}
\usepackage{graphicx}
\usepackage{balance}
\usepackage{calc}

\newtheorem{theorem}{Theorem}
\newtheorem{lemma}{Lemma}

\newcommand{\naive}{naive}

\input{preamble}

\begin{document}
\exclusivelicense

\setlength{\pdfpageheight}{\paperheight}
\setlength{\pdfpagewidth}{\paperwidth}

\conferenceinfo{ICFP~'13}{September 25--27, Boston, MA, USA.}
\copyrightyear{2013}
\copyrightdata{978-1-4503-2326-0/13/09}
\doi{2500365.2500604}


\title{Optimizing Abstract Abstract Machines}

\authorinfo{J. Ian Johnson}
           {Northeastern University}
           {ianj@ccs.neu.edu}
\authorinfo{Nicholas Labich}
           {Northeastern University}
           {labichn@ccs.neu.edu}
\authorinfo{Matthew Might}
           {University of Utah}
           {might@cs.utah.edu}
\authorinfo{David Van Horn}
           {Northeastern University}
           {dvanhorn@ccs.neu.edu}
\maketitle

\begin{abstract}
The technique of \emph{abstracting abstract machines} (AAM) provides a systematic approach for deriving computable approximations of evaluators that are easily proved sound.
This article contributes a complementary step-by-step process for subsequently going from a \naive{} analyzer derived under the AAM approach, to an efficient and correct implementation.
The end result of the process is a two to three order-of-magnitude improvement over the systematically derived analyzer, making it competitive with hand-optimized implementations that compute fundamentally less precise results.
\end{abstract}




\section{Introduction}

Program analysis provides sound predictive models of program behavior, but in order for such models to be effective, they must be efficiently computable and correct.
Past approaches to designing program analyses have often featured abstractions that are far removed from the original language semantics, requiring ingenuity in their construction and effort in their verification.
The \emph{abstracting abstract machines} (AAM) approach~\cite{dvanhorn:VanHorn2011Abstracting,dvanhorn:VanHorn2012Systematic} to deriving program analyses provides an alternative: a systematic way of transforming a programming language semantics in the form of an abstract machine into a family of abstract interpreters.
It thus reduces the burden of constructing and verifying the soundness of an abstract interpreter.
%

%

By taking a machine-oriented view of computation, AAM makes it possible to design, verify, and implement program analyzers for realistic language features typically considered difficult to model.
The approach was originally applied to features such as higher-order functions, stack inspection, exceptions, laziness, first-class continuations, and garbage collection.
It has since been used to verify actor-\cite{local:DOsualdo:12A} and thread-based~\cite{dvanhorn:Might2011Family} parallelism and behavioral contracts~\cite{dvanhorn:TobinHochstadt2012Higherorder};
it has been used to model Coq~\cite{local:harvard}, Dalvik~\cite{local:dalvik}, Erlang~\cite{local:DOsualdo:12B}, JavaScript~\cite{local:DBLP:journals/corr/abs-1109-4467}, and Racket~\cite{dvanhorn:TobinHochstadt2012Higherorder}.

\begin{figure}[t]
\small
\begin{center}
\includegraphics[width=3.2in]{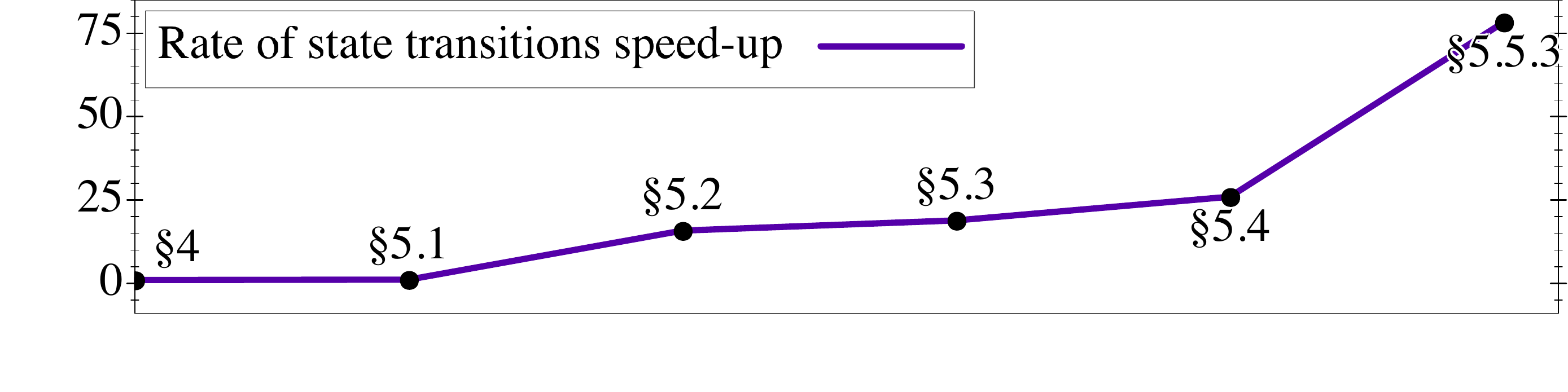}
\includegraphics[width=3.2in]{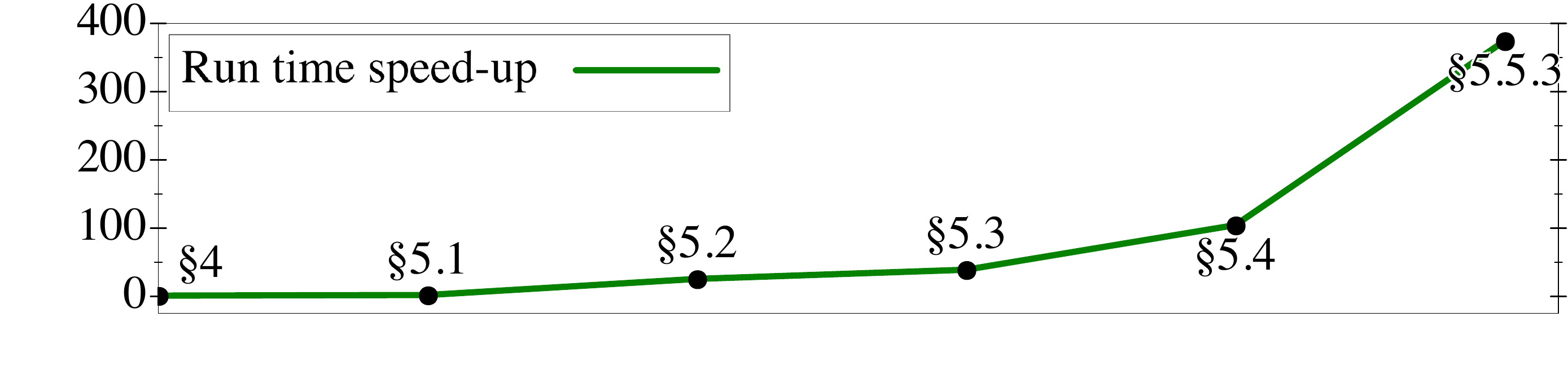}
\vspace{-1.5em}
\end{center}
\caption{
Factor improvements over the baseline analyzer for the \Church{} benchmark in terms of the rate of state transitions and total analysis time.
(Bigger is better.)
Each point is marked with the section that introduces the optimization.}
\label{fig:churchtime}
\end{figure}

The primary strength of the approach is that abstract interpreters can be easily derived through a small number of steps from existing machine models.
Since the relationships between abstract machines and higher-level semantic models---such as definitional interpreters~\cite{dvanhorn:reynolds-hosc98}, structured operational semantics~\cite{dvanhorn:Plotkin1981Structural}, and reduction semantics~\cite{dvanhorn:Felleisen2009Semantics}---are well understood~\cite{dvanhorn:Danvy:DSc}, it is possible to navigate from these high-level semantic models to sound program analyzers in a systematic way.
Moreover, since these analyses so closely resemble a language's interpreter
(a) implementing an analysis requires little more than implementing an interpreter,
(b) a single implementation can serve as both an interpreter and analyzer, and
(c) verifying the correctness of the implementation is straightforward.
Unfortunately, the AAM approach yields analyzers with poor performance relative to hand-optimized analyzers.
Our work takes aim squarely at this ``efficiency gap,'' and narrows it in an equally systematic way through a number of simple steps, many of which are inspired by run-time implementation techniques such as laziness and compilation to avoid interpretative overhead.
Each of these steps is proven correct, so the end result is an implementation that is trustworthy and efficient.
In this article, we develop a systematic approach to deriving a practical implementation of an abstract-machine-based analyzer using mostly semantic means rather than tricky engineering.
Our goal is to empower programming language implementers and researchers to explore and convincingly exhibit their ideas with a low barrier to entry.
The optimizations we describe are widely applicable and apparently effective to scale far beyond the size of programs typically considered in the recent literature on flow analysis for functional languages.

\section{At a glance}

We start with a quick review of the AAM approach to develop an analysis framework and then apply our step-by-step optimization techniques in the simplified setting of a core functional language.
This allows us to explicate the optimizations with a minimal amount of inessential technical overhead.
Following that, we scale this approach up to an analyzer for a realistic untyped, higher-order imperative language with a number of interesting features and then measure improvements across a suite of benchmarks.
At each step during the initial presentation and development, we evaluated the implementation on a set of benchmarks.
The highlighted benchmark in figure \ref{fig:churchtime} is from Vardoulakis and Shivers~\cite{dvanhorn:Vardoulakis2011CFA2} that tests distributivity of multiplication over addition on Church numerals.
For the step-by-step development, this benchmark is particularly informative:
\begin{enumerate}
\item it can be written in most modern programming languages,
\item it was designed to stress an analyzer's ability to deal with complicated environment and control structure arising from the use of higher-order functions to encode arithmetic, and
\item its improvement is about median in the benchmark suite considered in section~\ref{sec:eval}, and thus it serves as a good sanity check for each of the optimization techniques considered.
\end{enumerate}

\begin{figure}[t]
\small
\begin{center}
\begin{tabular}{ccc}
\raisebox{1ex-\height}{
\includegraphics[height=3.5in]{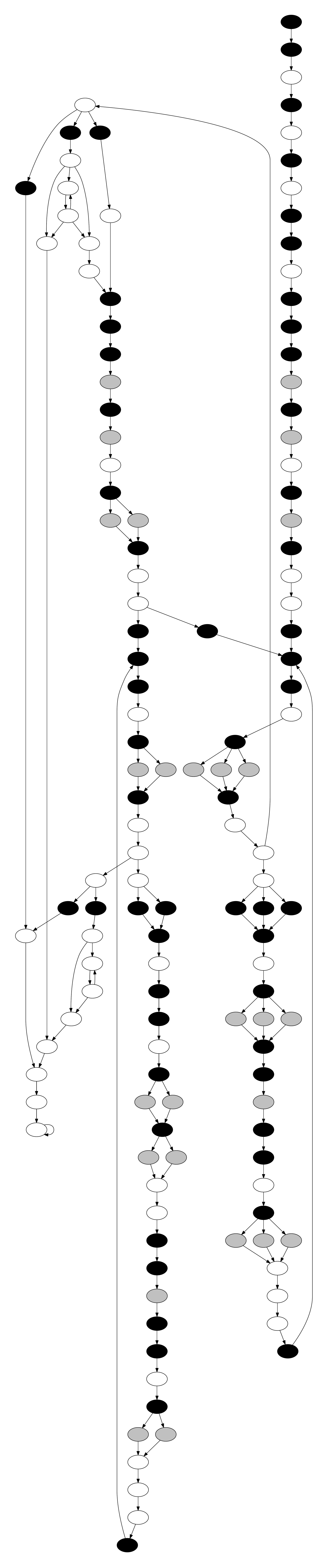}}
&
\raisebox{1ex-\height}{
\includegraphics[height=3.5in]{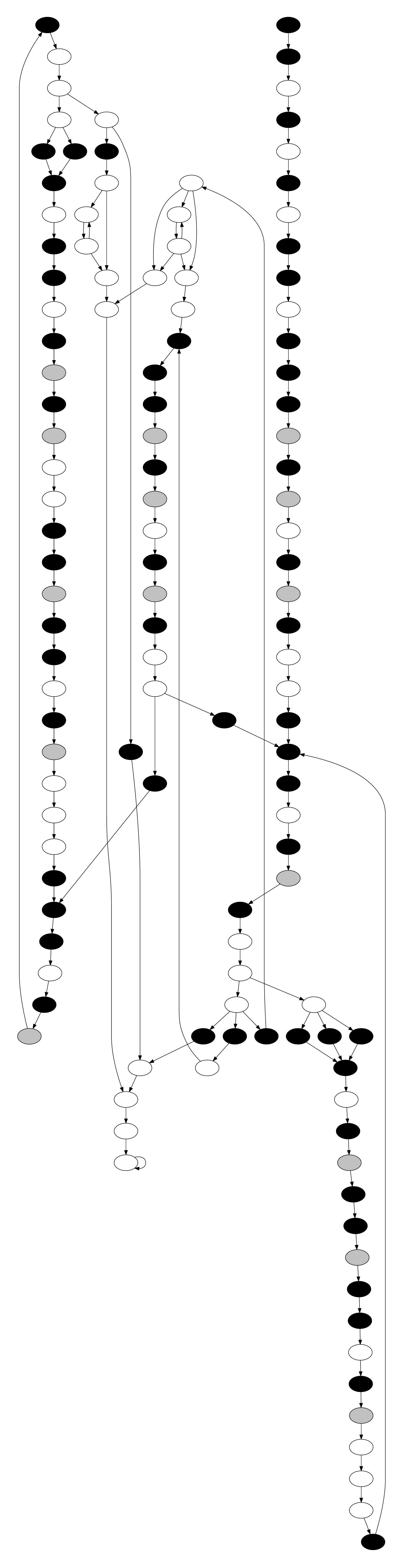}}
&
\raisebox{1ex-\height}{
\includegraphics[height=3in]{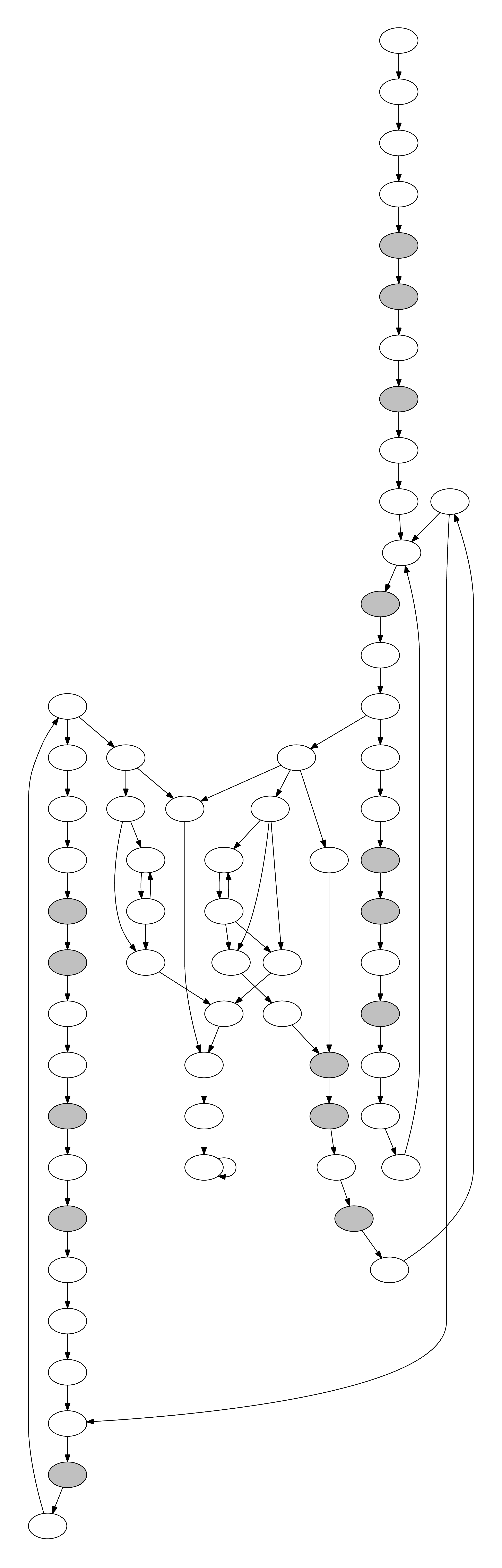}}
\\
(a) Baseline
&
(b) Lazy
&
(c) Compiled (\& lazy)
\end{tabular}
\end{center}
\caption{
Example state graphs for Earl et. al. program.
  Gray states follow variable references, {\tt ev} states are black, and all others are white.
 Part (a) shows the baseline analyzer result.
  It has long ``corridor'' transitions and ``diamond'' subgraphs that fan-out from nondeterminism and fan-in from joins.
  Part (b) shows the result of performing nondeterminism lazily and thus avoids many of the diamond subgraphs.
  Part (c) shows the result of abstract compilation that removes interpretive overhead in the form of intermediate states, thus minimizing the corridor transitions.
  The end result is a more compact abstraction of the program that can be generated faster.}
\label{fig:state-graphs}
\end{figure}

We start, in section~\ref{sec:aam}, by developing an abstract interpreter according to the AAM approach.
In the initial abstraction, each state carries a store (what is called per-state store variance).
The space of stores is exponential in size; without further abstraction, the analysis is exponential and thus cannot analyze the example in a reasonable amount of time.
In section~\ref{sec:baseline}, we perform a further abstraction by widening the store.
The resulting analyzer sacrifices precision for speed and is able to analyze the example in about 1 minute.
This step is described by Van Horn and Might~\cite[\S 3.5--6]{dvanhorn:VanHorn2012Systematic} and is necessary to make even small examples feasible.
We therefore take a widened interpreter as the baseline for our evaluation.
Section~\ref{sec:opt} gives a series of simple abstractions and implementation techniques that, in total, speed up the analysis by nearly a factor of 500, dropping the analysis time to a fraction of a second.
Figure~\ref{fig:churchtime} shows the step-wise improvement of the analysis time for this example.
The AAM approach, in essence, does the following: it takes a machine-based view of computation and turns it into a \emph{finitary approximation} by bounding the size of the store.
With a limited address space, the store must map addresses to \emph{sets} of values.
Store updates are interpreted as joins, and store dereferences are interpreted by non-deterministic choice of an element from a set.
The result of analyzing a program is a finite directed graph where nodes in the graph are (abstract) machine states and edges denote machine transitions between states.
The techniques we propose for optimizing analysis fall into the following categories:
\begin{enumerate}
\item generate fewer states by avoiding the eager exploration of non-deterministic choices that will later collapse into a single join point.
  We accomplish this by applying lazy evaluation techniques so that non-determinism is evaluated \emph{by need}.

\item generate fewer states by avoiding unnecessary, intermediate states of a computation.
  We accomplish this by applying compilation techniques from functional languages to avoid interpretive overhead in the machine transition system.

\item generate states faster.
  We accomplish this by better algorithm design in the fixed-point computation we use to generate state graphs.
\end{enumerate}
Figure~\ref{fig:state-graphs} shows the effect of (1) and (2) for the small motivating example in Earl, et al.~\cite{dvanhorn:Earl2012Introspective}.
By generating significantly fewer states at a significantly faster rate, we are able to achieve large performance improvements in terms of both time and space.
Section~\ref{sec:eval} describes the evaluation of each optimization technique applied to an implementation supporting a more realistic set of features, including mutation, first-class control, compound data, a full numeric tower and many more forms of primitive data and operations.
We evaluate this implementation against a set of benchmark programs drawn from the literature.
For all benchmarks, the optimized analyzer outperforms the baseline by at least a factor of
two to
three orders of magnitude.

Section~\ref{sec:related} relates this work to the literature and section~\ref{sec:conclusion} concludes.

\section{Abstract interpretation of ISWIM}
\label{sec:aam}

In this section, we give a brief review of the AAM approach by
defining a sound analytic framework for a core higher-order functional
language: Landin's ISWIM~\cite{dvanhorn:Landin1966Next}.
In the subsequent sections, we will explore optimizations for the analyzer in this simplified setting, but scaling these techniques to realistic languages is straightforward and has been done for the analyzer evaluated in section~\ref{sec:eval}.
ISWIM is a family of programming languages parameterized by a set of base values and operations.
To make things concrete, we consider a member of the ISWIM family with integers, booleans, and a few operations.
Figure~\ref{fig:syntax} defines the syntax of ISWIM.
It includes variables, literals (either integers, booleans, or operations), $\lambda$-expressions for defining procedures, procedure applications, and conditionals.
Expressions carry a label, $\mlab$, which is drawn from an unspecified set and denotes the source location of the expression; labels are used to disambiguate distinct, but syntactically identical pieces of syntax.
We omit the label annotation in contexts where it is irrelevant.

\begin{figure}
\small
\[
\begin{array}{l@{\qquad}rcl}
\text{Expressions} & \mexp &=& \svar[^\mlab]\mvar\\
&&|& \slit[^\mlab]\mlit\\
&&|& \slam[^\mlab]\mvar\mexp\\
&&|& \sapp[^\mlab]\mexp\mexp \\
&&|& \sif[^\mlab]\mexp\mexp\mexp \\
\text{Variables}&\mvar &=& \syntax{x}\ |\ \syntax{y}\ |\ \dots\\
\text{Literals}&\mlit &=& \mnum\ |\ \mbln\ |\ \mop\\
\text{Integers}&\mnum &=& \syntax{0}\ |\ \syntax{1}\ |\ \syntax{-1}\ |\ \dots\\
\text{Booleans}&\mbln &=& \strue\ |\ \sfalse\\
\text{Operations}&\mop &=& \syntax{zero?}\ |\ \syntax{add1}\ |\ \syntax{sub1}\ |\ \dots
\end{array}
\]
\caption{Syntax of ISWIM}
\label{fig:syntax}
\end{figure}

\begin{figure}
\small
\[
\begin{array}{l@{\qquad}rcl}
\text{Values} & \mval,\mvalx{u} &=& \clos{\mvar,\mexp,\menv}\ |\ \mlit\ |\ \mkont\\
\text{States} & \mstate &=& \ev[^\mcntr]{\mexp,\menv,\msto,\mkont}\\
                       &&|& \co{\mkont,\mval,\msto}\\
                       &&|& \ap[^\mcntr]{\mval,\mval,\msto,\mkont}\\
\text{Continuations} & \mkont &=& \kmt\\
&&|& \kfn{\mval,\mkaddr}\\
&&|& \kar{\mexp,\menv,\mkaddr}\\
&&|& \kif{\mexp,\mexp,\menv,\mkaddr}\\
\text{Addresses} &\maddr&\in&\Addr \\
\text{Times} &\mcntr&\in&\Counter \\
\text{Environments} &\menv&\in& \Var \rightharpoonup \Addr\\
\text{Stores} &\msto&\in& \Addr \rightharpoonup \wp(\Value)
\end{array}
\]
\caption{Abstract machine components}
\label{fig:domains}
\end{figure}

The semantics is defined in terms of a machine model.
The machine components are defined in figure~\ref{fig:domains};
figure~\ref{fig:aam} defines the transition relation (unmentioned components stay the same).
The evaluation of a program is defined as its set of traces that arise from iterating the machine transition relation.
The $\traces$ function produces the set of all proofs of reachability for any state $\mstate$ from the injection of program $\mexp$ (from which one could extract a string of states).
 The machine is a very slight variation on a standard abstract machine for ISWIM in ``eval, continue, apply'' form~\cite{dvanhorn:Danvy:DSc}.
It can be systematically derived from a definitional interpreter through a continuation-passing style transformation and defunctionalization, or from a structural operational semantics using the refocusing construction of Danvy and Nielsen~\cite{dvanhorn:Danvy-Nielsen:RS-04-26}.
Compared with the standard machine semantics, this definition is
different in the following ways, which make it abstractable as a
program analyzer:
\begin{itemize}
\item the store maps addresses to \emph{sets} of values,
\footnote{
More generally, we can have stores map to any domain that forms a Galois connection with sets of values, enabling $\interpdelta$ to produce elaborate abstractions of base values (e.g., interval or octagon abstractions).
We use sets of values for a simpler exposition.
}
not single values,

\item continuations are heap-allocated, not stack-allocated,
\item there are ``timestamps'' ($\mcntr \in \Counter$) and syntax labels ($\mlab$) threaded through the computation, and
\item the machine is implicitly parameterized by the functions $\alloc$, $\allockont$, $\tick$, $\interpdelta$, and spaces $\Addr$, $\Counter$ (and initial $\mcntr_0 \in \Counter$).
\end{itemize}

\begin{figure}
\small
\begin{gather*}
\begin{align*}
\traces(\mexp) &= \{ \ev[^{\mtcntr}]{\mexp,\varnothing,\varnothing,\kmt} \multimachstep \mstate \} \text{ where }
\end{align*}
\\[2mm]
\begin{array}{@{}r@{\ }c@{\ }l@{}}
\mstate &\machstep&\mstate' \text{ defined to be the following} \\
&&\text{let } \mcntr' =\tick(\mstate) \\
\ev{\svar\mvar,\menv,\msto,\mkont} &\machstep&
\co{\mkont,\mval,\msto}
\text{ if }\mval \in \msto(\menv(\mvar))
\\
\ev{\slit\mlit,\menv,\msto,\mkont} &\machstep&
\co{\mkont,\mlit,\msto}
\\
\ev[^\mcntr]{\slam\mvar\mexp,\menv,\msto,\mkont} &\machstep&
\co{\mkont,\clos{\mvar,\mexp,\menv},\msto}
\\
\ev[^\mcntr]{\sapp[^\mlab]{\mexpi0}{\mexpi1},\menv,\msto,\mkont} &\machstep&
\ev[^{\mcntr'}]{\mexpi{0},\menv,\msto',\kar[_\mlab^\mcntr]{\mexpi{1},\menv,\mkaddr}}
\\
&&
\text{ where }\mkaddr = \allockont^\mcntr_\mlab(\msto,\mkont) \\
&&\phantom{\text{ where }}\msto' = \msto\sqcup[\mkaddr \mapsto \setof{\mkont}]
\\
\ev[^\mcntr]{\sif[^\mlab]{\mexpi0}{\mexpi1}{\mexpi2},\menv,\msto,\mkont} &\machstep&
\ev[^{\mcntr'}]{\mexpi0,\menv,\msto',\kif[^\mcntr]{\mexpi1,\mexpi2,\menv,\mkaddr}}
\\
&&
\text{ where }\mkaddr = \allockont^\mcntr_\mlab(\msto,\mkont) \\
&&\phantom{\text{ where }}\msto' = \msto\sqcup[\mkaddr \mapsto \setof{\mkont}]
\\[2mm]
\co{\kar[^\mcntr_\mlab]{\mexp,\menv,\mkaddr},\mval,\msto} & \machstep&
\ev[^\mcntr]{\mexp,\menv,\msto,\kfn[^\mcntr_\mlab]{\mval,\mkaddr}}
\\
\co{\kfn[^\mcntr_\mlab]{{\mvalx{u}},\mkaddr},\mval,\msto} & \machstep&
\ap[^\mcntr_\mlab]{\mvalx{u},\mval,\mkont,\msto}
\text{ if }\mkont \in \msto(\mkaddr)
\\
\co{\kif[^\mcntr]{\mexpi0,\mexpi1,\menv,\mkaddr},\strue,\msto} & \machstep&
\ev[^{\mcntr'}]{\mexpi0,\menv,\msto,\mkont}
\text{ if }\mkont\in\msto(\mkaddr)
\\
\co{\kif[^\mcntr]{\mexpi0,\mexpi1,\menv,\mkaddr},\sfalse,\msto} & \machstep&
\ev[^{\mcntr'}]{\mexpi1,\menv,\msto,\mkont}
\text{ if }\mkont\in\msto(\mkaddr)
\\[2mm]
\ap[^\mcntr_\mlab]{\clos{\mvar,\mexp,\menv},\mval,\msto,\mkont} & \machstep&
\ev[^{\mcntr'}\!]{\mexp,\menv',\msto',\mkont}
\\
&&\text{ where } \maddr  =\alloc(\mstate) \\
&&\phantom{\text{ where }} \menv' = \menv[\mvar\mapsto\maddr] \\
&&\phantom{\text{ where }} \msto' = \msto\sqcup[\maddr\mapsto\{\mval\}] \\
\\
\ap[^\mcntr_\mlab]{\mop,\mval,\msto,\mkont} & \machstep&
\co{\mkont,\mval',\msto}
\text{ if } \mval'\in\interpdelta(\mop,\mval)
\end{array}
\end{gather*}
\caption{Abstract abstract machine for ISWIM}
\label{fig:aam}
\end{figure}

\paragraph{Concrete interpretation} To characterize concrete interpretation, set the implicit
parameters of the relation given in figure~\ref{fig:aam} as follows:
\begin{align*}
\alloc(\mstate) &= \maddr \mbox{ where } \maddr \notin \text{ the } \msto \text{ within } \mstate \\
\allockont^\mcntr_\mlab(\msto,\mkont) &=\mkaddr \mbox{ where } \mkaddr \notin \msto
\end{align*}
These functions appear to ignore $\mlab$ and $\mcntr$, but they can be
used to determinize the choice of fresh addresses.
The $\sqcup$ on stores in the figure is a point-wise lifting of $\cup$:
 $\msto \sqcup \msto' = \lambda \maddr. \msto(\maddr) \cup \msto'(\maddr)$.
The resulting relation is non-deterministic in its choice of addresses, however it must always choose a fresh address when allocating a continuation or variable binding.
If we consider machine states equivalent up to consistent renaming and fix an allocation scheme, this relation defines a deterministic machine (the relation is really a function).
The interpretation of primitive operations is defined by setting $\interpdelta$ as follows:
\begin{align*}
\mnum+1 &\in \interpdelta(\saddone,\mnum) &
\mnum-1 &\in \interpdelta(\ssubone,\mnum)\\
\strue &\in \interpdelta(\szerohuh,\szero) &
\sfalse &\in \interpdelta(\szerohuh,\mnum)\text{ if }\mnum\neq \szero\\
\end{align*}

\paragraph{Abstract interpretation}
To characterize abstract interpretation, set the implicit parameters just as above, but drop the $\maddr \not\in \msto$ condition.
The $\interpdelta$ relation takes some care to not make the analysis run forever; a simple instantiation is a flat abstraction where arithmetic operations return an abstract top element $\sNum$, and $\szerohuh$ returns both $\strue$ and $\sfalse$ on $\sNum$.
This family of interpreters is also non-deterministic in choices of addresses, but it is free to choose addresses that are already in use.
Consequently, the machines may be non-deterministic when multiple values reside in a store location.
It is important to recognize from this definition that \emph{any} allocation strategy is a sound abstract interpretation~\cite{dvanhorn:Might2009Posteriori}.
In particular, concrete interpretation is a kind of abstract interpretation.
So is an interpretation that allocates a single cell into which all bindings and continuations are stored.
The former is an abstract interpretation with uncomputable reachability and gives only the ground truth of a program's behavior; the latter is an abstract interpretation that is easy to compute but gives little information.
Useful program analyses lay somewhere in between and can be characterized by their choice of address representation and allocation strategy.
Uniform \(k\)-CFA~\cite{dvanhorn:nielson-nielson-popl97}, presented next, is one such analysis.
\paragraph{Uniform \(k\)-CFA}
To characterize uniform \(k\)-CFA, set the allocation strategy as follows, for a fixed constant \(k\):
{\small
\begin{align*}
\Counter &= \Label^* \\
\mtcntr &= \epsilon \\
\alloc(\ap[^\mcntr_\mlab]{\clos{\mvar,\mexp,\menv},\mval,\msto,\mkont}) &= \mvar\kpush[_k]{\mlab\mcntr} \\
\allockont^\mcntr_\mlab(\msto,\mkont) &= \mlab\mcntr \\
\tick(\ev[^\mcntr]{\mexp,\menv,\msto,\mkont}) &= \mcntr \\
\tick(\co{\kar[^\mcntr]{\mexp,\menv,\mkaddr},\mval,\msto}) &= \mcntr \\
\tick(\ap[^\mcntr_\mlab]{\mvalx{u},\mval,\mkont}) &= \kpush[_k]{\mlab\mcntr} \\
  \kpush[_0]{\mcntr} &= \kpush[_k]{\epsilon} = \mtcntr \\
  \kpush[_{k+1}]{\mlab\mcntr} &= \mlab\kpush[_k]{\mcntr} \\
\end{align*}}
The \(\lfloor\cdot\rfloor_k\) notation denotes the truncation of a list
of symbols to the leftmost \(k\) symbols.

All that remains is the interpretation of primitives.  For abstract
interpretation, we set $\interpdelta$ to the function that returns
$\sNum$ on all inputs---a symbolic value we interpret as denoting the
set of all integers.

At this point, we have abstracted the original machine to one which
has a finite state space for any given program, and thus forms the
basis of a sound, computable program analyzer for ISWIM.

\section{From machine semantics to baseline analyzer}
\label{sec:baseline}

The uniform $k$-CFA allocation strategy would make $\traces$ in figure
\ref{fig:aam} a computable abstraction of possible executions, but one
that is too inefficient to run, even on small examples.  Through this
section, we explain a succession of approximations to reach a more
appropriate baseline analysis.
We ground this path by first formulating the analysis in terms of a
classic fixed-point computation.

\subsection{Static analysis as fixed-point computation}
\label{sec:fixpoint}

Conceptually, the AAM approach calls for computing an analysis as a
graph exploration: (1) start with an initial state, and (2) compute
the transitive closure of the transition relation from that state. All
visited states are potentially reachable in the concrete, and all
paths through the graph are possible traces of execution.

We can cast this exploration process in terms of a fixed-point calculation.
Given the initial state $\mstate_0$ and the transition relation $\machstep$,
we define the global transfer function:
\begin{equation*}
 F_{\mstate_0} : \wp(\State) \times \wp(\State\times\State) \to \wp(\State) \times \wp(\State\times\State)\text.
\end{equation*}
Internally, this global transfer function computes the successors of all supplied states, and then includes the initial state:
\begin{align*}
  F_{\mstate_0}(V,E) &= (\{ \mstate_0 \} \cup V', E') \\
    E' &= \setof{ (\mstate,\mstate') \mid \mstate \in V \text{ and } \mstate \machstep \mstate'} \\
    V' &= \setof{ \mstate' \mid (\mstate,\mstate') \in E'}
\end{align*}
Then, the evaluator for the analysis computes the least fixed-point of the global transfer function:
 $\eval(\mexp) = \mathrm{lfp}(F_{\mstate_0})\text{,}$
where $\mstate_0 = \ev[^\mtcntr]{\mexp, \varnothing, \varnothing, \kmt}$.

The possible traces of execution tell us the most about a program, so we take $\traces(\mexp)$ to be the (regular) set of paths through the computed graph.
 We elide the construction of the set of edges in this paper.
To conduct this \naive{} exploration on the \Church{} example would require considerable time.
  Even though the state space is finite, it is exponential in the size of the program.
  Even with $k = 0$, there are exponentially many stores in the AAM framework.
In the next subsection, we fix this with store widening to reach polynomial (albeit of high degree) complexity.
This widening effectively lifts the store out of individual states to create a single, global shared store for all.

\subsection{Store widening}
\label{sec:storewiden}

A common technique to accelerate convergence in flow analyses is to share a common, global store.
Formally, we can cast this optimization as a second abstraction or as the application of a widening operator
\footnote{Technically, we would have to copy the value of the global store to all states being stepped to fit the formal definition of a widening, but this representation is order-isomorphic to that.}
during the fixed-point iteration.
In the ISWIM language, such a widening makes 0-CFA quartic in the size of the program.
Thus, complexity drops from intractable exponentiality to a merely daunting polynomial.
Since we can cast this optimization as a widening, there is no need to change the transition relation itself.
Rather, what changes is the structure of the fixed-point iteration.
In each pass, the algorithm will collect all newly produced stores and join them together.
Then, before each transition, it installs this joined store into current state.
To describe this process, AAM defined a transformation of the reduction relation so that it operates on a pair of a set of contexts ($C$) and a store ($\sigma$).
A context includes all non-store components, \emph{e.g.}, the expression, the environment and the stack.
The transformed relation, $\widehat{\machstep}$, is
{\small
\begin{align*}
(C, \msto) &\mathrel{\widehat{\machstep}} (C', \msto'), \\
\mbox{where } C' &= \{c' \mid \wn(c, \msto) \mathrel{\machstep} \wn(c', \msto^c), c \in C\} \\
              \msto' &= \bigsqcup\; \{\msto^c \mid \wn(c,\msto)\mathrel{\machstep} \wn(c', \msto^c), c \in C\} \\
\wn &: \Context \times \Store \to \State \\
\wn(\ev{\mexp, \menv, \mkont}, \msto) &= \ev{\mexp, \menv, \msto, \mkont} \\
\wn(\co{\mval, \mkont}, \msto) &= \co{\mval, \mkont, \msto} \\
\wn(\ap{\mvalx{u}, \mval, \mkont}, \msto) &= \ap{\mvalx{u}, \mval, \msto, \mkont} \\
\end{align*}}
To retain soundness, this store grows monotonically as the least upper bound of all occurring stores.
%

\subsection{Store-allocate all values}
\label{sec:baselineeval}

The final approximation we make to get to our baseline is to store-allocate all values that appear, so that any non-machine state that contains a value instead contains an address to a value.
The AAM approach stops at the previous optimization.
However, the \kfnalone continuation stores a value, and this makes the space of continuations quadratic rather than linear in the size of the program, for a monovariant analysis like 0-CFA.
Having the space of continuations grow linearly with the size of the program will drop the overall complexity to cubic (as expected).
We also need to allocate an address for the argument position in an \apalone state.
To achieve this linearity for continuations, we allocate an address for the value position when we create the continuation.
This address and the tail address are both determined by the label of the application point, so the space becomes linear and the overall complexity drops to cubic.
This is a critical abstraction in languages with $n$-ary functions, since otherwise the continuation space grows super-exponentially (${\mathcal O}(n^n)$).
We extend the semantics to additionally allocate an address for the function value when creating the $\kfnalone$ continuation.
The continuation has to contain this address to remember where to retrieve values from in the store.
The new evaluation rules follow, where $\mcntr' = \tick(\mstate)$:
\begin{align*}
\co[^\mcntr]{\kar{\mexp,\menv,\mkaddr},\mval,\msto} & \machstep
\ev[^{\mcntr'}]{\mexp,\menv,\msto',\kfn{\maddr,\mkaddr}} \\
\text{ where }
  \maddr &= \alloc(\mstate) \\
  \msto' &= \ext{\msto}{\maddr}{\{\mval\}}
\end{align*}
Now instead of storing the evaluated function in the continuation frame itself, we indirect it through the store for further control on complexity and precision:
\begin{align*}
\co[^\mcntr]{\kfn{\maddr,\mkaddr},\mval,\msto} & \machstep
\ap[^{\mcntr'}_\mlab]{\mvalx{u},\maddr,\mkont,\msto'}
\\
\text{ if } \mkont &\in \msto(\mkaddr), \mvalx{u} \in \msto(\maddr) \\
\text{ where } \maddr &= \alloc(\mstate) \\
               \msto' &= \ext{\msto}{\maddr}{\{\mval\}}
\end{align*}
Associated with this indirection, we now apply all functions stored in the address.
This nondeterminism is necessary in order to continue with evaluation.



\section{Implementation techniques}
\label{sec:opt}

In this section, we discuss the optimizations for abstract interpreters that
yield our ultimate performance gains.
We have two broad categories of these optimizations: (1) pragmatic
improvement, (2) transition elimination.
The pragmatic improvements reduce overhead and trade space for time
by utilizing:
\begin{enumerate}
 \item timestamped stores;
 \item store deltas; and
 \item imperative, pre-allocated data structures.
\end{enumerate}
The transition-elimination optimizations reduce the overall number of transitions
made by the analyzer by performing:
\begin{enumerate}
  \setcounter{enumi}{3}
 \item frontier-based semantics;
 \item lazy non-determinism; and
 \item abstract compilation.
\end{enumerate}
All pragmatic improvements are precision preserving (form complete abstractions), but the semantic changes are not in some cases, for reasons we will describe.
 We did not observe the precision differences in our evaluation.
We apply the frontier-based semantics combined with timestamped stores as our first step.
  The move to the imperative will be made last in order to show the effectiveness of these techniques in the purely functional realm.

\subsection{Timestamped frontier}

The semantics given for store widening in section \ref{sec:storewiden}, while simple, is wasteful.
 It also does not model what typical implementations do.
 It causes all states found so far to step each iteration, even if they are not revisited.
 This has negative performance \emph{and} precision consequences (changes to the store can travel back in time in straight-line code).
 We instead use a frontier-based semantics that corresponds to the classic worklist algorithms for analysis.
 The difference is that the store is not modified in-place, but updated after all frontier states have been processed.
 This has implications for the analysis' precision and determinism.
 Specifically, higher precision, and it is deterministic even if set iteration is not.
The state space changes from a store and set of contexts to a set of seen abstract states (context plus store), $\mseen$, a set of contexts to step (the frontier), $F$, and a store to step those contexts with, $\msto$:
\begin{equation*}
(\mseen, F, \msto) \mathrel{\widehat{\machstep}} (\mseen \cup \mseen', F', \msto')
\end{equation*}

We constantly see more states, so $\mseen$ is always growing. The frontier,
which is what remains to be done, changes. Let's start with the result
of stepping all the contexts in $F$ paired with the current store (call it $I$ for intermediate):
\begin{equation*}
I = \setof{(c',\msto') \mid \wn(c, \msto) \mathrel{\machstep} \wn(c', \msto'), c \in F}
\end{equation*}

The next store is the least upper bound of all the stores in $I$:

\begin{equation*}
\msto' = \bigsqcup\; \setof{\msto \mid (\_,\msto) \in I}
\end{equation*}
The next frontier is exactly the states that we found from stepping
the last frontier, but have not seen before. They must be
\emph{states}, so we pair the contexts with the next store:
\begin{equation*}
F' = \setof{c \mid (c,\_) \in I, (c,\msto') \notin \mseen}
\end{equation*}
Finally, we add what we know we had not yet seen to the seen set:
\begin{equation*}
  \mseen' = \setof{(c,\msto') \mid c \in F'}
\end{equation*}
To inject a program $e$ into this machine, we start off knowing we have seen the first state, and that we need to process the first state:
\begin{align*}
\inject(e) &= (\setof{\ttuple{c_0}{\bot}},\setof{c_0},\bot) \\
\text{ where } c_0 &= \ev{e,\bot,\kmt}
\end{align*}
Notice that now $\mseen$ has several copies of the abstract store in it.
As it is, this semantics is much less efficient (but still more precise) than the previously proposed semantics because membership checks have to compare entire stores.
Checking equality is expensive because the stores within each state are large, and nearly every entry must be checked against every other due to high similarities amongst stores.
And, there is a better way.
Shivers' original work on $k$-CFA was susceptible to the same problem, and he suggested three complementary optimizations:
 (1) make the store global;
 (2) update the store imperatively; and
 (3) associate every change in the store with a version number -- its timestamp.
Then, put timestamps in states where previously there were stores.
Given two states, the analysis can now compare their stores just by comparing their timestamps -- a constant-time operation.

There are two subtle losses of precision in Shivers' original timestamp technique that we can fix.

\begin{enumerate}
\item{In our semantics, the store does not change until the entire frontier has been explored.
 This avoids cross-branch pollution which would otherwise happen in Shivers' semantics, e.g., when one branch writes to address $\maddr$ and another branch reads from address $\maddr$.}
\item{The common implementation strategy for timestamps destructively updates each state's timestamp.
 This loses \emph{temporal} information about the contexts a state is visited in, and in what order.
 Our semantics has a drop-in replacement of timestamps for stores in the seen set ($\mseentime$), so we do not experience precision loss.}
\end{enumerate}

{\small
\begin{align*}
\Sigma \in \Store^* \qquad \mseentime \subseteq {\mathbb N} \times \Context \qquad F \subseteq \Context
\end{align*}
\begin{align*}
(\mseentime, F, \msto, \Sigma, t) &\mathrel{\widehat{\machstep}^T} (\mseentime \cup \mseentime', F', \msto', \Sigma', t') \\
\mbox{where } I &= \setof{(c',\msto^c) \mid \wn(c, \msto) \mathrel{\machstep} \wn(c', \msto^c), c \in F} \\
              \msto' &= \bigsqcup\; \{\msto^c \mid (\_,\msto^c) \in I\} \\
              (t',\Sigma') &=\left\{\begin{array}{ll}
                           (t+1,\msto'\Sigma') & \text{ if } \msto' \neq \msto \\
                           (t,\Sigma)   & \text{ otherwise}
                          \end{array}\right. \\
              F' &= \setof{c \mid (c,\_) \in I, (c,t') \notin \mseentime} \\
              \mseentime' &= \setof{(c,t') \mid c \in F'}\\
\inject(e) &= (\setof{\ttuple{c_0}{0}},\setof{c_0},\bot,\cons{\bot}{\epsilon},0) \\
\text{ where } c_0 &= \ev{e,\bot,\kmt}
\end{align*}}

The observation Shivers made was that the store is increasing monotonically, so all stores throughout execution will be totally ordered (form a chain).
 This observation allows you to replace stores with pointers into this chain.
 We keep the stores around in $\Sigma$ to achieve a complete abstraction.
 This corresponds to the temporal information about the execution's effect on the store.

Note also that $F$ is only populated with states that have not been seen at the resulting store.
 This is what produces the more precise abstraction than the baseline widening.

The general fixed-point combinator we showed above can be specialized to this semantics, as well.
 In fact, $\widehat{\machstep}^T$ is a functional relation, so we can get the least fixed-point of it directly.

\begin{lemma}
  $\widehat{\machstep}$ maintains the invariant that all stores in $\mseen$ are totally ordered and $\msto$ is an upper bound of the stores in $\mseen$.
\end{lemma}

\begin{lemma}
  $\widehat{\machstep}^T$ maintains the invariant that $\Sigma$ is in order with respect to $\sqsupset$ and $\msto = \hd(\Sigma)$.
\end{lemma}

\begin{theorem}
$\widehat{\machstep}^T$ is a complete abstraction of $\widehat{\machstep}$.
\end{theorem}
The proof follows from the order isomorphism that, in one direction, sorts all the stores in $\mseen$ to form $\Sigma$, and translates stores in $\mseen$ to their distance from the end of $\Sigma$ (their timestamp).
 In the other direction, timestamps in $\mseentime$ are replaced by the stores they point to in $\Sigma$.

\subsection{Locally log-based store deltas}

The above technique requires joining entire (large) stores together.
 Additionally, there is still a comparison of stores, which we established is expensive.
 Not every step will modify all addresses of the store, so joining entire stores is wasteful in terms of memory and time.
 We can instead log store changes and replay the change log on the full store after all steps have completed, noting when there is an actual change.
 This uses far fewer join and comparison operations, leading to less overhead, and is precision-preserving.

We represent change logs as $\msdiff \in \StoreDelta = (\Addr \times \Set(\Storeable))^*$.
 Each $\msto\sqcup[\maddr \mapsto \mval{s}]$ becomes a log addition $\cons{\ttuple{\maddr}{\mval{s}}}{\msdiff}$, where $\msdiff$ begins empty ($\mtlst$) for each step.
 Applying the changes to the full store is straightforward:
\begin{equation*}
\replay : (\StoreDelta \times \Store) \to (\Store \times \Boolean)
\end{equation*}
\begin{align*}
\replay(\left[ \ttuple{\maddr_i}{\mval{s_i}}, \ldots\right], \msto) &=
\ttuple{\msto'}
       {\diffp(\mval{s_i}, \msto(\maddr_i)) \vee \ldots} \\
\text{ where } \msto' &= {\msto \sqcup [\maddr_i \mapsto \mval{s_i}] \sqcup \ldots} \\
\diffp(\mval{s}, \mval{s'}) &= \mval{s'} \overset{?}{=} \mval{s} \sqcup \mval{s'}
\end{align*}

We change the semantics slightly to add to the change log rather than produce an entire modified store.
The transition relation is identical except for the addition of this change log.
  We maintain the invariant that lookups will never rely on the change log, so we can use the originally supplied store unmodified.

A taste of the changes to the reduction relation is as follows:
\begin{equation*}
\dmachstep \subseteq (\Context\times\Store) \times (\Context\times\StoreDelta) 
\end{equation*}
\begin{align*}
\ttuple{\ap[^\mcntr_\mlab]{\clos{\mvar,\mexp,\menv},\maddr,\mkont}}{\msto} & \dmachstep
\ttuple{\ev[^{\mcntr'}]{\mexp,\menv',\mkont}}{\cons{\ttuple{\maddr'}{\msto(\maddr)}}{\epsilon}} \\
\text{ where }\maddr' &= \alloc(\mstate) \\
              \menv' &= \menv[\mvar\mapsto\maddr']
\end{align*}


We lift $\dmachstep$ to accommodate for the asymmetry in the input and output, and change the frontier-based semantics in the following way:
{\small
\begin{align*}
(\mseentime, F, \msto,\Sigma,t) &\mathrel{\damachstep} (\mseentime \cup \mseentime', F', \msto',\Sigma',t') \\
\mbox{ where }
 I &= \setof{(c',\msdiff) \mid \ttuple{c}{\msto} \dmachstep \ttuple{c'}{\msdiff} } \\
 \ttuple{\msto'}{\changep} &= \replay(\appendall(\setof{\msdiff \mid (\_,\msdiff) \in I}),\msto) \\
 \ttuple{t'}{\Sigma'} &=
     \left\{
       \begin{array}{ll}
         \ttuple{t+1}{\msto\Sigma} & \text{ if } \changep \\
         \ttuple{t}{\Sigma} & \text{ otherwise}
       \end{array}\right. \\
 F' &= \setof{c \mid (c,\_) \in I, (c,t') \notin \mseentime} \\
 \mseentime' &= \setof{(c,t') \mid c \in F'} \\
\appendall(\varnothing) &= \epsilon \\
\appendall(\setof{\msdiff}\cup\Xi) &= \append(\msdiff,\appendall(\Xi))
\end{align*}}
Here $\appendall$ combines change logs across all non-deterministic steps for a state to later be replayed.
The order the combination happens in doesn't matter, because join is associative and commutative.
\begin{lemma}
$\ttuple{c}{\msto} \dmachstep \ttuple{c'}{\msdiff}$ iff $\wn(c,\msto) \machstep \wn(c',\replay(\msdiff,\msto))$
\end{lemma}
By cases on $\dmachstep$ and $\machstep$.
\begin{lemma}[$\changep$ means change]
Let $\replay(\msdiff,\msto) = \ttuple{\msto'}{\changep}$. $\msto' \neq \msto$ iff $\changep$.
\end{lemma}
By induction on $\msdiff$.
\begin{theorem}
$\damachstep$ is a complete abstraction of $\widehat{\machstep}^T$.
\end{theorem}
Follows from previous lemma and that join is associative and commutative.




\subsection{Lazy non-determinism}

Tracing the execution of the analysis reveals an immediate shortcoming: there is a high degree of branching and merging in the exploration.
Surveying this branching has no benefit for precision.
For example, in a function application, {\tt (f x y)}, where {\tt f}, {\tt x} and {\tt y} each have several values each argument evaluation induces $n$-way branching, only to be ultimately joined back together in their respective application positions.
Transition patterns of this shape litter the state-graph:
\vspace{-1em}
\begin{center}
\includegraphics[scale=0.2]{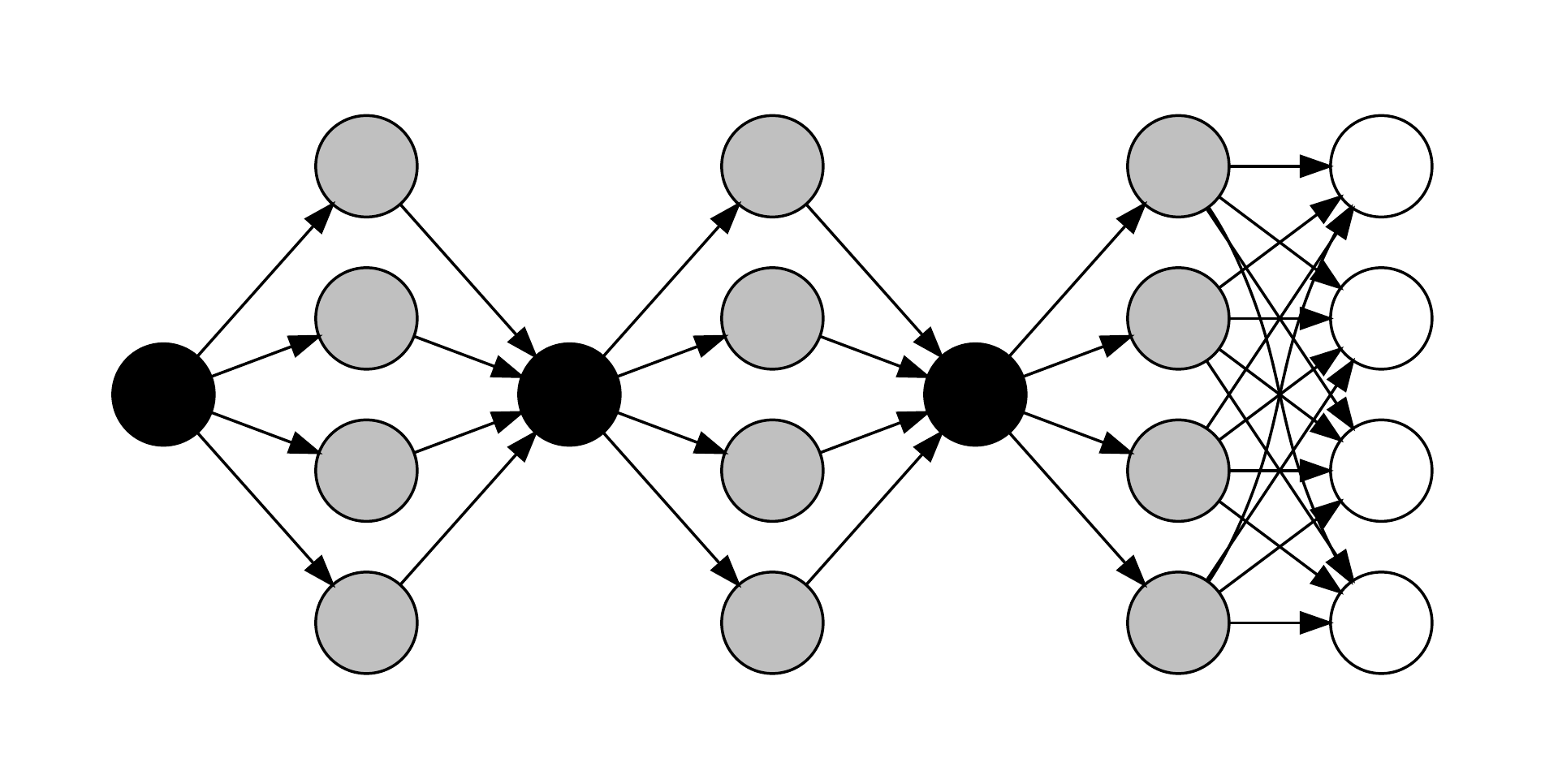}
\end{center}
\vspace{-1em}
To avoid the spurious forking and joining, we {\it delay} the non-determinism until and unless it is needed in {\it strict contexts} (such as the guard of an {\tt if}, a called procedure, or a numerical primitive application).
Doing so collapses these forks and joins into a linear sequence of states:
\vspace{-1em}
\begin{center}
\includegraphics[scale=0.2]{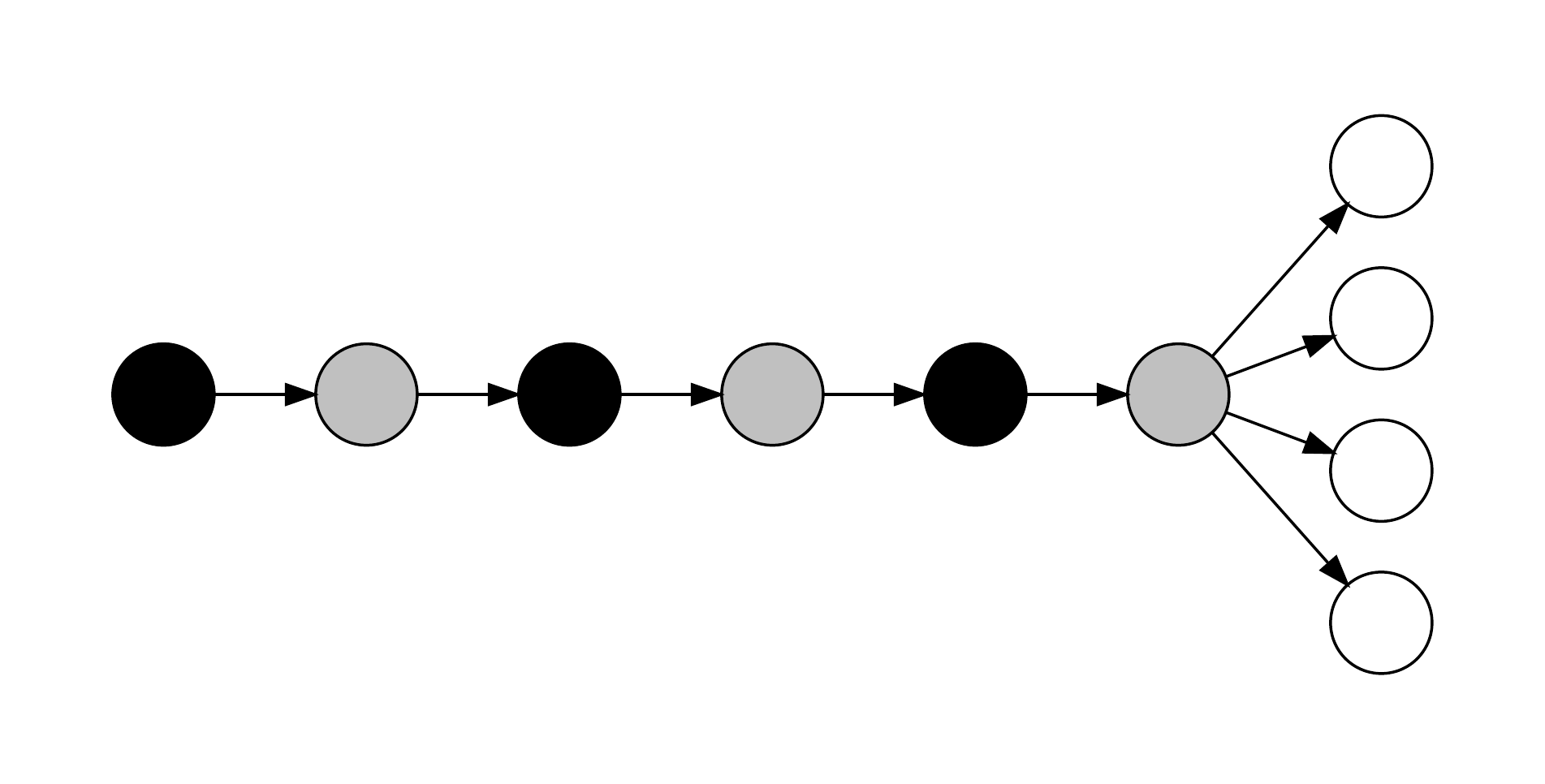}
\end{center}
\vspace{-1em}
This shift does not change the concrete semantics of the language to be lazy.
Rather, it abstracts over transitions that the original non-deterministic semantics steps through.
We say the abstraction is \emph{lazy} because it delays splitting on the values in an address until they are \emph{needed} in the semantics.
It does not change the execution order that leads to the values that are stored in the address.

We introduce a new kind of value,
\spchoice
{$\saddr{\maddr}$}
{$\superposition{\mval{s}}$ (for ``superposition'')},
that represents a delayed non-deterministic choice of a value from
\spchoice
{$\msto(\maddr)$}
{$\mval{s}$}.
The following rules highlight the changes to the semantics:

\renewcommand{\ext}{\mathit{ext}}
{\small
\begin{align*}
\spchoice
{\force &: \Store \times \Value \to \Set(\Value) \\
 \force(\msto,\saddr{\maddr}) &= \msto(\maddr) \\
 \force(\msto,\mval) &= \setof{\mval}}
{\force &: \Value \to \Set(\Value) \\
 \force(\superposition{\mval{s}}) &= \mval{s} \\
 \force(\mval) &= \{\mval\}}
\\
\ev{\svar{\mvar},\menv,\mkont,\msto} &\lmachstep\;
\spchoice
{\co{\mkont,\saddr{\menv(\mvar)},\msto}}
{\co{\mkont,\superposition{\msto(\menv(\mvar))},\msto}} \\
\co{\kar[^\mcntr_\mlab]{\mexp,\menv,\mkaddr},\mval,\msto}
&\lmachstep\;
\ev[^{\mcntr'}]{\mexp,\menv,\msto',\kfn[^\mcntr_\mlab]{\maddr_f,\mkaddr}} \\
\text{ where }
\maddr_f &= \alloc(\mstate) \\
\msto' &=
\spchoice
{\msto \sqcup[\maddr \mapsto \force(\msto,\mval)]}
{\msto \sqcup[\maddr \mapsto \force(\mval)]} \\
\co{\kif[^\mcntr]{\mexpi0,\mexpi1,\menv,\mkaddr},\mval,\msto}
&\lmachstep\;
\ev[^{\mcntr'}]{\mexpi0,\menv,\msto,\mkont} \\
\text{ if } \mkont &\in \msto(\mkaddr),
            \strue \in \spchoice{\force(\msto,\mval)}{\force(\mval)}
\end{align*}}
Since {\tt if} guards are in strict position, we must force the value to determine which branch to to take.
The middle rule uses $\force$ only to combine with values in the store - it does not introduce needless non-determinism.
\spchoice{
\noindent
We have two choices for how to implement lazy non-determinism.

\paragraph{Option 1: Lose precision; simplify implementation}
This semantics introduces a subtle precision difference over the baseline.
Consider a configuration where a reference to a variable and a binding of a variable will happen in one step, since store widening leads to stepping several states in one big ``step.''
With laziness, the reference will mean the original binding(s) of the variable \emph{or} the new binding, because the actual store lookup is delayed one step (i.e. laziness is administrative).

\paragraph{Option 2: Regain precision; complicate implementation}
The administrative nature of laziness means that we could remove the loss in precision by storing the result of the lookup in a value representing a delayed nondeterministic choice.
This is a more common choice in 0CFA implementations we have seen, but it interferes with the next optimization due to the invariant from store deltas we have that lookups must not depend on the change log.

}{}

\spchoice
{
\begin{theorem}[Soundness]
  If $\mstate \machstep \mstate'$ and $\mstate \sqsubseteq \mastate$ then there exists a $\mastate'$ such that $\mastate \lmachstep \mastate'$ and $\mstate' \sqsubseteq \mastate'$
\end{theorem}
Here $\sqsubseteq$ is straightforward --- the left-hand side store must be contained in the right-hand-side store, and if values occur in the states, the left-hand-side value must be in the forced corresponding right-hand-side value.
The proof is by cases on $\mstate \machstep \mstate'$.}
{
\begin{theorem}[Completeness]For all $\mexp$,
 $\traces_{\text{Lazy}}(\mexp)$ is a complete abstraction of $\traces_{\text{ISWIM}}(\mexp)$.
\end{theorem}
We have a statement about traces because we need induction to show no cruft values are in superposition.
The induction hypothesis tells us that there are non-lazy traces that lead to all the values in superposition, so when we take a lazy step, we are taking several non-lazy steps, and we stay in sync.
 The other direction we just collapse the superposition in each possibility to construct the non-lazy traces.}


\subsection{Abstract compilation}

The prior optimization saved time by doing the same amount of reasoning as before but in fewer transitions.
 We can exploit the same idea---same reasoning, fewer transitions---with abstract compilation.
 Abstract compilation transforms complex expressions whose \emph{abstract} evaluation is deterministic into ``abstract bytecodes.''
  The abstract interpreter then does in one transition what previously took many.
  Refer back to figure \ref{fig:state-graphs} to see the effect of abstract compilation.
 In short, abstract compilation eliminates unnecessary allocation, deallocation and branching.
 The technique is precision preserving without store widening.
 We discuss the precision differences with store widening at the end of the section.

The compilation step converts expressions into functions that expect the other components of the {\tt ev} state.
Its definition in figure \ref{fig:compile} shows close similarity to the rules for interpreting {\tt ev} states.
 The next step is to change reduction rules that create {\tt ev} states to instead call these functions.
 Figure \ref{fig:caam} shows the modified reduction relation.
 The only change from the previous semantics is that $\evalone$ state construction is replaced by calling the compiled expression.
 For notational coherence, we write $\lambda^\mcntr(\mathit{args} \ldots)$ for $\lambda(\mathit{args} \ldots, \mcntr)$ and $\mcomp^\mcntr(\mathit{args}\ldots)$ for $\mcomp(\mathit{args}\ldots,
\mcntr)$.

\begin{figure}
\small
\begin{align*}
\compile{\_} &: \Expr \to \Store \\
             &\phantom{: \Expr } \to \Env  \times \StoreDelta \times \Kont \times \Counter \\
             &\phantom{: \Expr } \to \State \\
\mcntr' &= \tick(\mlab,\menv,\msto,\mcntr) \\
\compile{\svar\mvar}_\msto &=
 \lambda^\mcntr(\menv,\msdiff,\mkont) .
\spchoice
{\co{\mkont,\saddr{\menv(\mvar)}},\msdiff}
{\co{\mkont,\superposition{\msto(\menv(\mvar))}},\msdiff}
\\
\compile{\slit\mlit}_\msto &= \lambda^\mcntr(\menv,\msdiff,\mkont) .
\co{\mkont,\mlit},\msdiff
\\
\compile{\slam\mvar\mexp}_\msto &= \lambda^\mcntr(\menv,\msdiff,\mkont) .
\co{\mkont,\clos{\mvar,\compile{\mexp},\menv}},\msdiff
\\
\compile{\sapp[^\mlab]{\mexpi0}{\mexpi1}}_\msto &= \lambda^\mcntr(\menv,\msdiff,\mkont) .
\compile{\mexpi0}^{\mcntr'}(\menv,\msdiff',\kar[_\mlab^\mcntr]{\compile{\mexpi1},\menv,\mkaddr})
\\
&\setlength\arraycolsep{5pt}
\begin{array}{lrl}
\text{ where } & \mkaddr = \allockont^\mcntr_\mlab(\msto,\mkont) \\
               & \msdiff' = \cons{\ttuple{\mkaddr}{\setof{\mkont}}}{\msdiff}
\end{array}
\\
\compile{\sif[^\mlab]{\mexpi0}{\mexpi1}{\mexpi2}}_\msto &= \lambda^\mcntr(\menv,\msdiff,\mkont) .
\compile{\mexpi0}^{\mcntr'}(\menv,\msdiff',\kif[^\mcntr]{\compile{\mexpi1},\compile{\mexpi2},\menv,\mkaddr})
\\
&\text{ where }\mkaddr = \allockont^\mcntr_\mlab(\msto,\mkont) \\
&\phantom{\text{ where }} \msdiff' = \cons{\ttuple{\mkaddr}{\setof{\mkont}}}{\msdiff}
\end{align*}
\caption{Abstract compilation}
\label{fig:compile}
\end{figure}

\begin{figure}
\small
\begin{gather*}
\begin{align*}
\traces(\mexp) &= \setof{ \inject(\compile{\mexp}^\mtcntr_\bot(\bot,\epsilon,\kmt)) \multimachstep \mstate}
                    \text{ where } \\
\inject(c,\msdiff) &= \wn(c,\replay(\msdiff,\bot)) \\
\wn(c,\msto) \machstep \wn(c',\msto') &\iff c \cmachstep_\msto c',\msdiff \\
\msdiff \text{ is such that } &\replay(\msdiff,\msto) = \msto'
\end{align*}
\\[2mm]
\begin{align*}
\co{\kar[^\mcntr_\mlab]{\mcomp,\menv,\mkaddr},\mval} & \cmachstep_\msto
\mcomp^\mcntr(\msto)(\menv,\msdiff,\kfn[^\mcntr_\mlab]{\maddr_f,\mkaddr}) \\
\text{ where } \maddr_f &= \alloc(\mstate) \\
               \msdiff &= \cons{\ttuple{\maddr_f}{\force(\msto,\mval)}}{\epsilon}
\\
\co{\kfn[^\mcntr_\mlab]{\maddr_f,\mkaddr},\mval} & \cmachstep_\msto
\ap[^\mcntr_\mlab]{\mvalx{u},\maddr,\mkont},\cons{\ttuple{\maddr}{\force(\msto,\mval)}}{\epsilon} \\
\text{ if }\mvalx{u} &\in \msto(\maddr_f), \mkont \in \msto(\mkaddr)
\\
\co{\kif[^\mcntr]{\mcompi0,\mcompi1,\menv,\mkaddr},\strue} & \cmachstep_\msto
\mcompi{0}^\mcntr(\msto)(\menv,\epsilon,\mkont)
\text{ if }\mkont\in\msto(\mkaddr)
\\
\co{\kif[^\mcntr]{\mcompi0,\mcompi1,\menv,\mkaddr},\sfalse} & \cmachstep_\msto
\mcompi{1}^\mcntr(\msto)(\menv,\epsilon,\mkont)
\text{ if }\mkont\in\msto(\mkaddr)
\\[2mm]
\ap[^\mcntr_\mlab]{\clos{\mvar,\mcomp,\menv},\maddr,\mkont} & \cmachstep_\msto
\mcomp^{\mcntr'}(\msto)(\menv',\msdiff,\mkont) \\
\text{ where }\menv' &= \menv[\mvar\mapsto\maddr] \\
              \msdiff &= \cons{\ttuple{\maddr}{\msto(\maddr)}}{\epsilon}
\\
\ap{\mop,\maddr,\mkont} & \cmachstep_\msto
\co{\mkont,\mvalx{u}},\epsilon \\
\text{ where } \mval &\in \spchoice{\msto(\maddr)}{\msto(\maddr)}, \mvalx{u}\in\interpdelta(\mop,\mval)
\end{align*}
\end{gather*}
\caption{Abstract abstract machine for compiled ISWIM}
\label{fig:caam}
\end{figure}

\paragraph{Correctness}
The correctness of abstract compilation seems obvious, but it has never before been rigorously proved.
What constitutes correctness in the case of dropped states, anyway?
Applying an abstract bytecode's function does many ``steps'' in one go, at the end of which, the two semantics line up again (modulo representation of expressions).
 This constitutes the use of a notion of stuttering.
 We provide a formal analysis of abstract compilation \emph{without} store widening with a proof of a stuttering bisimulation~\cite{ianjohnson:BCG88} between this semantics and lazy non-determinism without widening to show precision preservation.

The number of transitions that can occur in succession from an abstract bytecode is roughly bounded by the amount of expression nesting in the program.
We can use the expression containment order to prove stuttering bisimulation with a well-founded equivalence bisimulation (WEB)~\cite{ianjohnson:manolios-diss}.
%
 WEBs are equivalent to the notion of a stuttering bisimulation, but are more amenable to mechanization since they also only require reasoning over one step of the reduction relation.
The trick is in defining a well-founded ordering that determines when the two semantics will match up again, what Manolios calls the pair of functions $\erankt$ and $\erankl$ (but we don't need $\erankl$ since the uncompiled semantics doesn't stutter).
We define a refinement, $r$, from non-compiled to compiled states (built structurally) by ``committing'' all the actions of an $\evalone$ state (defined similarly to $\compile{\_}$, but immediately applies the functions), and subsequently changing all expressions with their compiled variants.
 Since WEBs are for single transition systems, a WEB refinement is over the disjoint union of our two semantics, and the equivalence relation we use is just that a state is related to its refined state (and itself).
 Call this relation $B$.
Before we prove this setup is indeed a WEB, we need one lemma that applying an abstract bytecode's function is equal to refining the corresponding $\evalone$ state:
\begin{lemma}[Compile/commit]
Let $c,\msdiff' = \compile{\mexp}^\mcntr_{r(\msto)}(\menv,\msdiff,r(\mkont))$.
Let $\wn(c',\msto') = r(\ev[^\mcntr]{\mexp,\menv,\msto,\mkont})$.
$\wn(c,\replay(\msdiff',\msto)) = \wn(c',\replay(\msdiff,\msto'))$.
\end{lemma}
The proof is by induction on $\mexp$.

\begin{theorem}[Precision preservation]
$B$ is a WEB on $\lmachstep \uplus \machstep$
\end{theorem}

The proof follows by cases on $\lmachstep \uplus \machstep$ with the WEB \emph{witness} being the well-order on expressions (with a $\bot$ element), and the following $\erankt$, $\erankl$ functions:

\begin{align*}
\erankt(\ev[^\mcntr]{\mexp,\menv,\msto,\mkont}) &= \mexp \\
\erankt(\mstate) &= \bot \quad \text{otherwise} \\
\erankl(s,s') &= 0
\end{align*}
All cases are either simple steps or appeals to the well-order on $\erankt$'s range.
The other rank function, $\erankl$ is unnecessary, so we just make it the constant 0 function.
The $\cmachstep$ cases are trivial.

\paragraph{Wide store and abstract compilation}
%
 It is possible for different stores to occur between the different semantics because abstract compilation can change the order in which the store is changed (across steps).
 This is the case because some ``corridor'' expressions may compile down to change the store before some others, meaning there is no stuttering relationship with the wide lazy semantics.
%
%
%
 Although there is a difference pre- and post- abstract compilation, the result is still deterministic in contrast to Shivers' technique.
 The soundness is in tact since we can add store-widening the correct unwidened semantics with an easy correctness proof.
 Call $\camachstep$ the result of the widening operator from the previous section on $\cmachstep$.





\subsection{Imperative, pre-allocated data structures}

Thus far, we have made our optimizations in a purely functional manner.
For the final push for performance, we need to dip into the imperative.
In this section, we show an alternative representation of the store and seen set that are more space-efficient and are amenable to destructive updates by adhering to a history for each address.
The following transfer function has several components that can be destructively updated, and intermediate sets can be elided by adding to global sets.
In fact, the log of store deltas can be removed as well, by updating the store in-place, and on lookup, using the first value timestamped $\le$ the current timestamp.
We start with the purely functional view.

\subsubsection{Pure setup for imperative implementation}

The store maps to a stack of timestamped sets of abstract values.
Throughout this section, we will be taking the parameter $t$ to be the ``current time,'' or the length of the store chain at the beginning of the step.
\vspace{-0.9em}
\begin{align*}
\msto \in \Store &= \Addr \to \Valstack \\
\mvalstack \in \Valstack &= (\Timestamp \times \wp(\Storeable))^*
\end{align*}

To allow imperative store updates, we maintain an invariant that we never look up values tagged at a time in the future:

{\small
\begin{align*}
\lookup(\mvalstack,t) &=
  \left\{
    \begin{array}{ll}
      \mval{s} & \text{ if } \mvalstack = \cons{\ttuple{t'}{\mval{s}}}{\mvalstack'}, t' \le t \\
      \mval{s'} & \text{ if } \mvalstack = \cons{\ttuple{t'}{\mval{s}}}{\cons{\ttuple{t''}{\mval{s'}}}{\mvalstack'}}, t' > t
    \end{array}\right.
\end{align*}}
To construct this value stack, we have a time-parameterized join operation that also tracks changes to the store.
If joining with a time in the future, we just add to it.
Otherwise, we're making a change for the future ($t+1$), but only if there is an actual change.
{\small
\begin{align*}
\msto \sqcup_t [\maddr \mapsto \mval{s}] &= \msto[\maddr \mapsto \mvalstack],\changep \\
\text{where } (\mvalstack,\changep) &= \msto(\maddr)\sqcup_t \mval{s} \\
\epsilon \sqcup_t \mval{s} &= \ttuple{t}{\mval{s}},\strue \\
\cons{\ttuple{t'}{\mval{s}}}{\mvalstack} \sqcup_t \mval{s'} &= \cons{\ttuple{t'}{\mval{s}\sqcup\mval{s'}}}{\mvalstack},\diffp(\mval{s}, \mval{s'}) \text{ if } t' > t \\
\mvalstack \sqcup_t \mval{s} &= \cons{\ttuple{t+1}{\mval{s^*}}}{\mvalstack},\strue
           \text{ if } \mval{s_t} \neq \mval{s^*} \\
 \text{where } &\mval{s_t} = \lookup(\msto(\maddr),t) \\
 &\mval{s^*} = \mval{s} \sqcup \mval{s_t} \\
\mvalstack \sqcup_t \mval{s} &= \mvalstack,\sfalse \text{ otherwise}
\end{align*}}

For the purposes of space, we reuse the $\cmachstep$ semantics, although the $\replay$ of the produced $\msdiff$ objects should be in-place, and the $\lookup$ function should be using this single-threaded store.
Because the store has all the temporal information baked into it, we rephrase the core semantics in terms of a transfer function.
The least fixed-point of this function gives a more compact representation of the reduction relation of the previous section.

{\small
\begin{align*}
\System &= (\widehat{\State} \to {\Timestamp}^*) \times \wp(\widehat{\State}) \times \Store \times \Timestamp \\
{\mathcal F} &: \System \to \System \\
{\mathcal F}(\mseentime,F, \msto,t) &= (\mseentime',F',\msto', t') \\
\text{ where }
I &= \setof{(c',\msdiff) \mid
       c \in F,
       c \cmachstep_{\msto^*} c',\msdiff} \\
\msto^* &= \lambda \maddr.\lookup(\msto(\maddr),t) \\
\ttuple{\msto'}{\changep} &= \replay(\appendall(\setof{\msdiff \mid (\_,\msdiff) \in I}),\msto) \\
t' &= \left\{\begin{array}{ll} t+1 & \text{ if } \changep \\
              t   & \text{ otherwise}
             \end{array}\right. \\
F' &= \setof{c \mid (c,\_) \in I, \changep \vee \mseentime(c) \neq \cons{t}{\_}} \\
\mseentime' &= \lambda c. \left\{\begin{array}{ll}
                               \cons{t'}{\mseentime(c)} & \text{ if } c \in F' \\
                               \mseentime(c) & \text{ otherwise}
                             \end{array}\right.
\end{align*}}

We prove semantic equivalence with the previous semantics with a
lock-step bisimulation with the stack of stores abstraction, which
follow from equational reasoning from the following lemmas:

\begin{lemma}
Stores of value stacks completely abstract stacks of stores.
\end{lemma}
This depends on some well-formedness conditions about the order of the
stacks. The store of value stacks can be translated to a stack of
stores by taking successive ``snapshots'' of the store at different
timestamps from the max timestamp it holds down to 0. Vice versa, we
replay the changes across adjacent stores in the stack.

We apply a similar construction to the different representation of seen states in order to get the final result:

\begin{theorem}
${\mathcal F}$ is a complete abstraction of $\camachstep$.
\end{theorem}

\subsubsection{Pure to imperative}

The intermediate data structures of the above transfer function can all be streamlined into globals that are destructively updated.
In particular, there are 5 globals:

\begin{enumerate}
\item{$\mseentime$: the \emph{seen} set, though made a map for faster membership tests and updates.}
\item{$F$: the \emph{frontier} set, which must be persisent or copied for the iteration through the set to be correct.}
\item{$\msto$: the store, which represents all stores that occur in the machine semantics.}
\item{$t$: the timestamp, or length of the store chain.}
\item{$\changep$: whether the store changed when stepping states in $F$.}
\end{enumerate}

The reduction relation would then instead of building store deltas, update the global store.
We would also not view it as a general relation, but a function that adds all next states to $F$ if they have
not already been seen.
At the end of iterating through $F$, $\mseentime$ is updated with the new states at the next timestamp.
There is no cross-step store poisoning since the lookup is restricted to the current step's time, which points to the same value throughout the step.

\subsubsection{Pre-allocating the store}

Internally, the algorithm at this stage uses hash tables to model the store to allow arbitrary address representations.
But, such a dynamic structure isn't necessary when we know the structure of the store in advance.
In a monovariant allocation strategy, the domain of the store is bounded by the number of expressions in the program.
If we label each expression with a unique natural, the analysis can index directly into the store without a hash or a collision.
Even for polyvariant analyses, it is possible to compute the maximum number of addresses and similarly pre-allocate either the spine of the store or (if memory is no concern) the entire store.

\section{Evaluation}
\label{sec:eval}

We have implemented, optimized, and evaluated an analysis framework supporting higher-order functions, state, first-class control, compound data, and a large number of primitive kinds of data and operations such as floating point, complex, and exact rational arithmetic.
The analysis is evaluated against a suite of Scheme benchmarks drawn from the literature.
\footnote{Source code of the implementation and benchmark suite available at \url{https://github.com/dvanhorn/oaam}}
For each benchmark, we collect analysis times, peak memory usage (as determined by Racket's GC statistics), and the rate of states-per-second explored by the analysis for each of the optimizations discussed in section~\ref{sec:opt}, cumulatively applied.
The analysis is stopped after consuming 30 minutes of time or 1 gigabyte of space
\footnote{All benchmarks are calculated as an average of 5 runs, done in parallel (each isolated to a core), on an 12-core, 64-bit Intel Xeon machine running at 2.40GHz with 12Gb of memory.}.
When presenting \emph{relative} numbers, we use the timeout limits as a lower bound on the actual time required (i.e., one minute versus timeout is at least 30 times faster), thus giving a conservative estimate of improvements.

%
%
For those benchmarks that did complete on the baseline, the optimized analyzer outperformed the baseline by a factor of two to three orders of magnitude.

We use the following set of benchmarks:
\begin{figure}
\centering
\include{bench-overview}
\caption{Overview performance comparison between baseline and
  optimized analyzer (entries of \text{{\small $t$}} mean timeout, and \text{{\small $m$}} mean out of memory).}
\label{fig:bench-overview}
\end{figure}

\begin{enumerate}  

\item {\bf nucleic}: a floating-point intensive application taken from molecular biology that has been used widely in benchmarking functional language implementations~\cite{dvanhorn:Hartel1996Benchmarking} and analyses (e.g.~\cite{dvanhorn:wright-jagannathan-toplas98,dvanhorn:jagannathan-etal-popl98}).
  It is a constraint satisfaction algorithm used to determine the three-dimensional structure of nucleic acids.

\item {\bf matrix} tests whether a matrix is maximal among all matrices of the same dimension obtainable by simple reordering of rows and columns and negation of any subset of rows and columns.
  It is written in continuation-passing style (used in \cite{dvanhorn:wright-jagannathan-toplas98,dvanhorn:jagannathan-etal-popl98}).

\item {\bf nbody}: implementation~\cite{ianjohnson:nbody87} of the Greengard multipole algorithm for computing gravitational forces on point masses distributed uniformly in a cube (used in \cite{dvanhorn:wright-jagannathan-toplas98,dvanhorn:jagannathan-etal-popl98}).
\item {\bf earley}: Earley's parsing algorithm, applied to a 15-symbol input according to a simple ambiguous grammar.
  A real program, applied to small data whose exponential behavior leads to a peak heap size of half a gigabyte or more during concrete execution.

\item {\bf maze}: generates a random maze using Scheme's {\tt call/cc} operation and finds a path solving the maze (used in \cite{dvanhorn:wright-jagannathan-toplas98,dvanhorn:jagannathan-etal-popl98}).

\item {\bf church}: tests distributivity of multiplication over addition for Church numerals (introduced by \cite{dvanhorn:Vardoulakis2011CFA2}).

\item {\bf lattice}: enumerates the order-preserving maps between two finite lattices (used in \cite{dvanhorn:wright-jagannathan-toplas98,dvanhorn:jagannathan-etal-popl98}).

\item {\bf boyer}: a term-rewriting theorem prover (used in \cite{dvanhorn:wright-jagannathan-toplas98,dvanhorn:jagannathan-etal-popl98}).

\item {\bf mbrotZ}: generates Mandelbrot fractal using complex numbers.

\item {\bf graphs}: counts the number of directed graphs with a distinguished root and \(k\) vertices, each having out-degree at most 2.
 It is written in a continuation-passing style and makes extensive use of higher-order procedures---it creates closures almost as often as it performs non-tail procedure calls (used by \cite{dvanhorn:wright-jagannathan-toplas98,dvanhorn:jagannathan-etal-popl98}).
\end{enumerate}

\begin{figure*}
\begin{center}
  \includegraphics[width=6.5in]{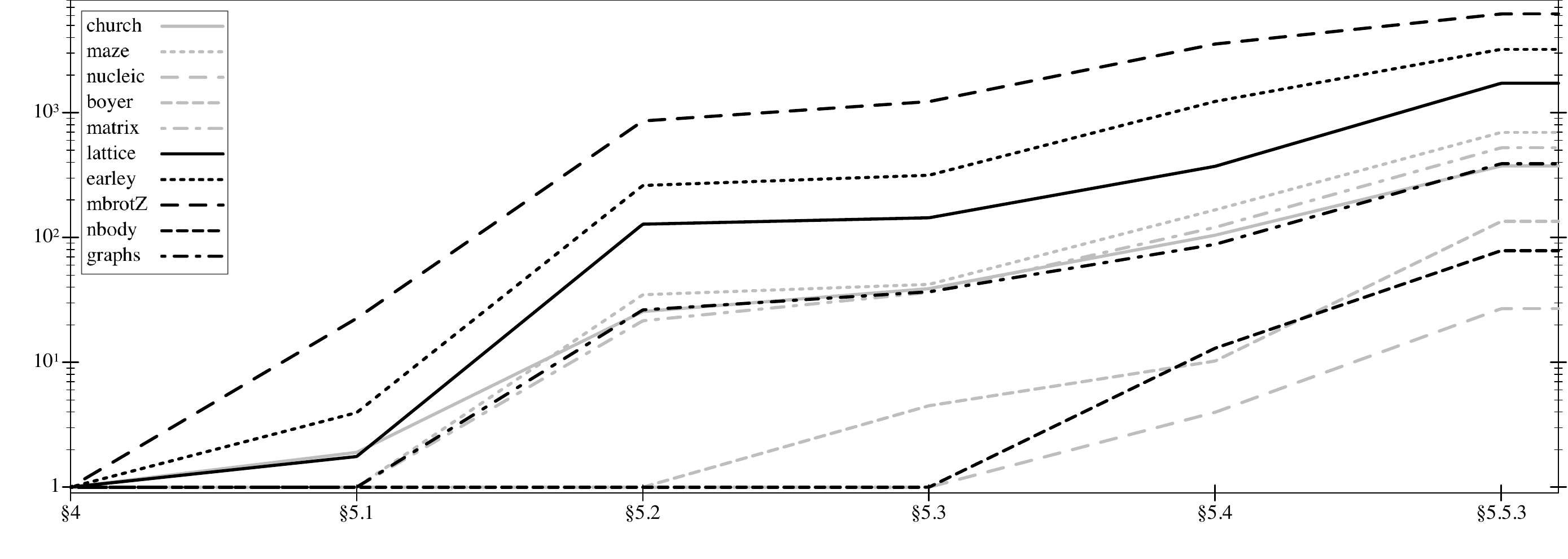}

  (a) Total analysis time speed-up (baseline / optimized)

  \vspace{1em}
  \includegraphics[width=6.5in]{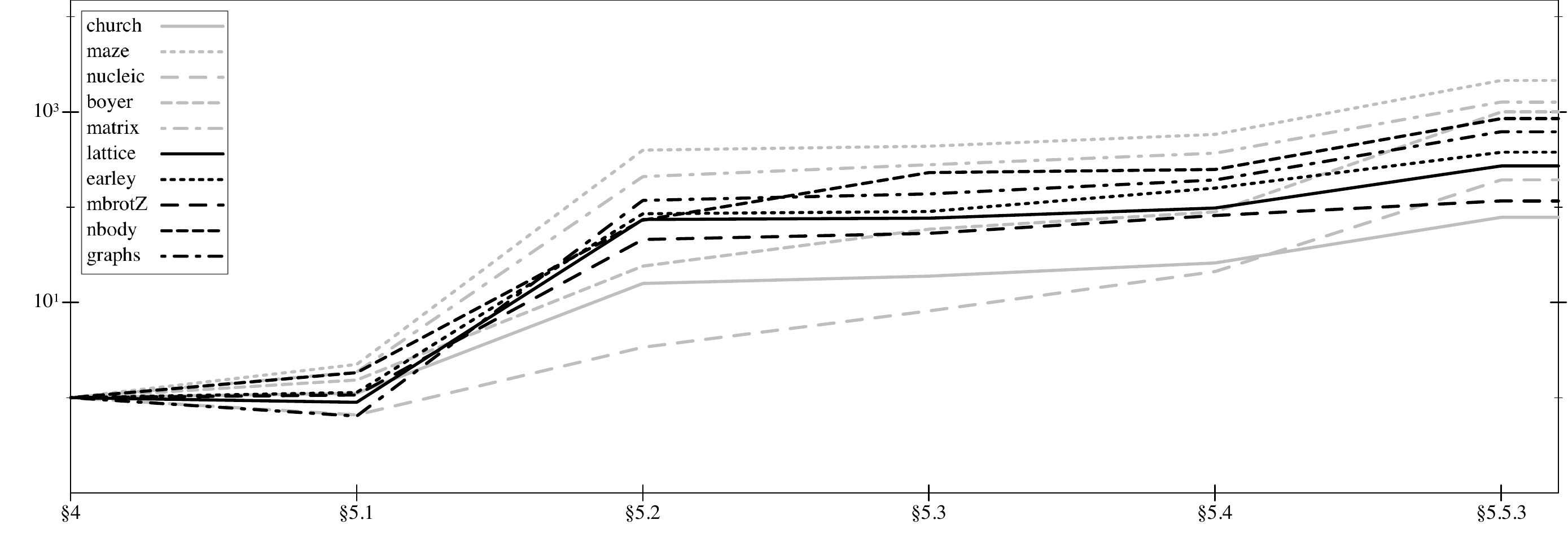}

  (b) Rate of state transitions speed-up (optimized / baseline)

  \vspace{1em}
  \includegraphics[width=6.5in]{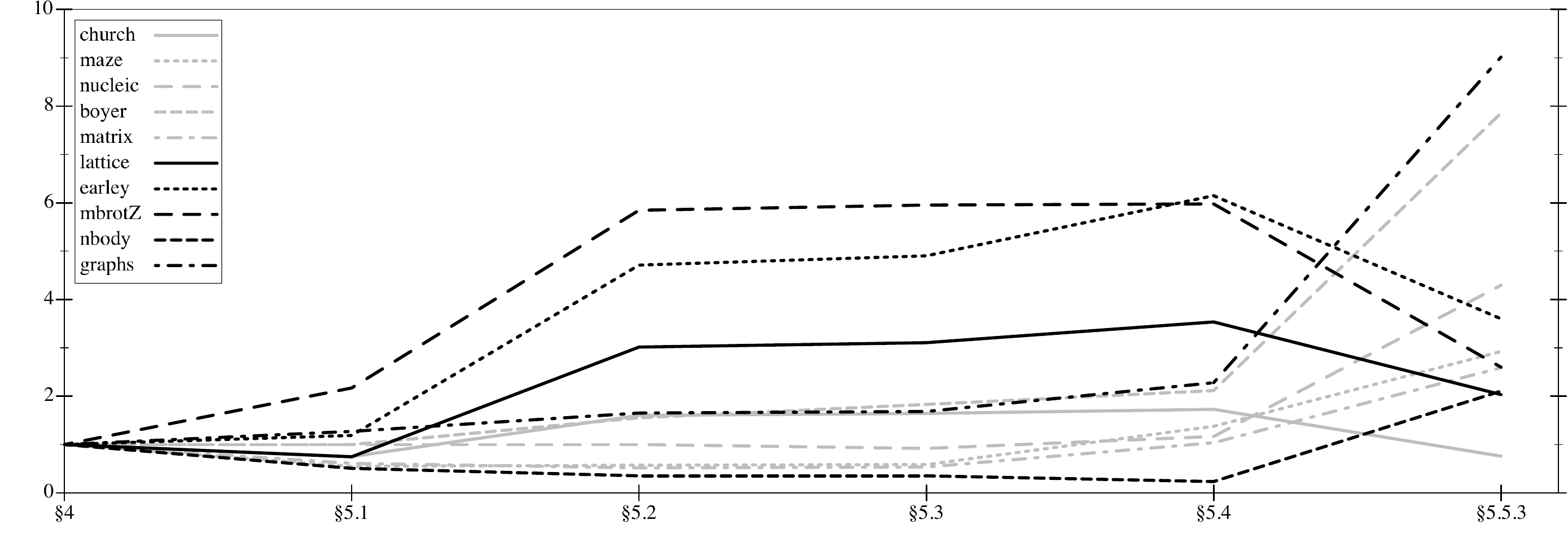}

  (c) Peak memory usage improvement (baseline / optimized)
\end{center}
\caption{Factors of improvement over baseline for each step of optimization (bigger is better).}
\label{fig:bench-all}
\end{figure*}

Figure~\ref{fig:bench-overview} gives an overview of the benchmark results in terms of absolute time, space, and speed between the baseline and most optimized analyzer.
Figure~\ref{fig:bench-all} plots the factors of improvement over the baseline for each optimization step.


To determine the impact of each section's technique on precision, we evaluated a singleton variable analysis to find opportunities to inline constants and closed functions.
We found no change in the results across all implementations, including Shivers' timestamp approximation---from an empirical point of view, these techniques are precision preserving despite the theoterical loss of precision.
Our step-wise optimizations strictly produce better analysis times with no observed loss of precision.
The final result is a systematically derived and verified implementation that operates within a small factor performance loss compared to a hand-optimized, unverified implementation.
Moreover, much of the performance gains are achieved with purely functional methods, which allow the use of these methods in rewriting tools and others with restricted input languages.
Peak memory usage is often considerably improved by the end of the optimization steps, but the effect of mutation has unexpected consequence on the analyzers memory consumption, likely due to a larger-than-necessary preallocated store.

\paragraph{Comparison with other flow analysis implementations}

The analysis considered here computes results similar to Earl, et al.'s 0-CFA implementation~\cite{dvanhorn:Earl2012Introspective}, which times out on the \Church{} benchmark because it does not widen the store as described for our baseline evaluator.
So even though it offers a fair point of comparison, a more thorough evaluation is probably uninformative as the other benchmarks are likely to timeout as well (and it would require significant effort to extend their implementation with the features needed to analyze our benchmark suite).
That implementation is evaluated against much smaller benchmarks: the largest program is 30 lines.
Vardoulakis and Shivers evaluate their CFA2 analyzer~\cite{dvanhorn:Vardoulakis2011CFA2} against a variant of 0-CFA defined in their framework and the example we draw on is the largest benchmark Vardoulakis and Shivers consider.
More work would be required to scale the analyzer to the set of features required by our benchmarks.
The only analyzer we were able to find that proved capable of analyzing the full suite of benchmarks considered here
was the Polymorphic splitting system of Wright and Jagannathan~\cite{dvanhorn:wright-jagannathan-toplas98}.
\footnote{This is not a coincidence; these papers set a high standard for evaluation, which we consciously aimed to approach.}
Unfortunately, these analyses compute an inherently different and incomparable form of analysis via a global acceptability judgment.
Consequently, we have omitted a complete comparison with these implementations.
The AAM approach provides more precision in terms of temporal-ordering of program states, which comes at a cost that can be avoided in constraint-based approaches.
Consequently implementation techniques cannot be ``ported'' between these two approaches.
However, our optimized implementation is within an order of magnitude of the performance of Wright and Jaganathan's analyzer.
Although we would like to improve this to be more competitive, the optimized AAM approach still has many strengths to recommend it in terms of precision, ease of implementation and verification, and rapid design.
 We can get closer to their performance by relying on the representation of addresses and the behavior of $\alloc$ to pre-allocate most data structures and split the abstract store out into parts that are more quickly accessed and updated.
Our semantic optimizations can still be applied to an analysis that does abstract garbage collection~\cite{dvanhorn:Might:2006:GammaCFA}, whereas the polymorphic splitting implementation is tied strongly to a single-threaded store.

\section{Related work}
\label{sec:related}

\paragraph{Abstracting Abstract Machines}

This work clearly closely follows Van Horn and Might's original papers on abstracting abstract machines~\cite{dvanhorn:VanHorn2011Abstracting,dvanhorn:VanHorn2012Systematic}, which in turn is one piece of the large body of research on flow analysis for higher-order languages (see Midtgaard~\cite{dvanhorn:Midtgaard2011Controlflow} for a thorough survey).
The AAM approach sits at the confluence of two major lines of research:
(1) the study of abstract machines~\cite{dvanhorn:landin-64} and their systematic construction~\cite{dvanhorn:reynolds-hosc98},
and (2) the theory of abstract interpretation \cite{dvanhorn:Cousot:1977:AI,dvanhorn:Cousot1979Systematic}.

\paragraph{Frameworks for flow analysis of higher-order programs}

Besides the original AAM work, the analysis most similar to that presented in section~\ref{sec:aam} is the infinitary control-flow analysis of Nielson and Nielson~\cite{dvanhorn:nielson-nielson-popl97} and the unified treatment of flow analysis by Jagannathan and Weeks~\cite{dvanhorn:jagannathan-weeks-popl95}.
Both are parameterized in such a way that in the limit, the analysis is equivalent to an interpreter for the language, just as is the case here.
What is different is that both give a constraint-based formulation of the abstract semantics rather than a finite machine model.

\paragraph{Abstract compilation}

Boucher and Feeley \cite{dvanhorn:Boucher1996Abstract} introduced the idea of abstract compilation, which used closure generation \cite{dvanhorn:Feeley1987Using} to improve the performance of control flow analysis.
We have adapted the closure generation technique from compositional evaluators to abstract machines and applied it to similar effect.

\paragraph{Constraint-based program analysis for higher-order languages}

Constraint-based program analyses (e.g.~\cite{dvanhorn:nielson-nielson-popl97,dvanhorn:wright-jagannathan-toplas98,dvanhorn:Meunier2006Modular,dvanhorn:steckler-wand-toplas97}) typically compute sets of abstract values for each program point.
These values approximate values arising at run-time for each program point.
Value sets are computed as the least solution to a set of (inclusion or equality) constraints.
The constraints must be designed and proved as a sound approximation of the semantics.
Efficient implementations of these kinds of analyses often take the form of worklist-based graph algorithms for constraint solving, and are thus quite different from the interpreter implementation.
The approach thus requires effort in constraint system design and implementation, and the resulting system require verification effort to prove the constraint system is sound and that the implementation is correct.
This effort increases substantially as the complexity of the analyzed language increases.
Both the work of maintaining the concrete semantics and constraint system (and the relations between them) must be scaled simultaneously.
However, constraint systems, which have been extensively studied in their own right, enjoy efficient implementation techniques and can be expressed in declarative logic languages that are heavily optimized~\cite{dvanhorn:bravenboer-smaragdakis-oopsla09}.
Consequently, constraint-based analyses can be computed quickly.
For example, Jagannathan and Wright's polymorphic splitting implementation~\cite{dvanhorn:wright-jagannathan-toplas98} analyses the \Church{} benchmark about 5.5 times faster than the fastest implementation considered here.
These analyses compute very different things, so the performance comparison is not apples-to-apples.
The AAM approach, and the state transition graphs it generates, encodes temporal properties not found in classical constraint-based analyses for higher-order programs.
Such analyses (ultimately) compute judgments on program terms and contexts, e.g., at expression $e$, variable $x$ may have value $v$.
The judgments do not relate the order in which expressions and context may be evaluated in a program, e.g., it has nothing to say with regard to question like,
``Do we always evaluate $e_1$ before $e_2$?''
%
%
The state transition graphs can answer these kinds of queries, but evaluation demonstrated this does not come for free.

\section{Conclusion}
\label{sec:conclusion}

Abstract machines are not only a good model for rapid analysis development, they can be systematically developed into efficient algorithms that can be proved correct.
We view the primary contribution of this work as a systematic path that eases the design, verification, and implementation of analyses using the abstracting abstract machine approach to within a factor of performant constraint-based analyses.


\paragraph{Acknowledgments}

We thank Suresh Jagannathan for providing source code to the
polymorphic splitting
analyzer~\cite{dvanhorn:wright-jagannathan-toplas98} and Ilya Sergey
for the introspective pushdown
analyzer~\cite{dvanhorn:Earl2012Introspective}.  
We thank Sam Tobin-Hochstadt for encouragement and feedback---he was
the first to prompt us to look into how to make effective
implementations of the AAM approach.
We thank Vincent St-Amour and Mitchell Wand for feedback on early drafts
and Greg Morrisett and Matthias Felleisen for discussions.
We thank our anonymous reviewers for their detailed comments.
This material is based on research sponsored by DARPA under the
programs Automated Program Analysis for Cybersecurity
(FA8750-12-2-0106) and Clean-Slate Resilient Adaptive Hosts (CRASH).
The U.S. Government is authorized to reproduce and distribute reprints
for Governmental purposes notwithstanding any copyright notation
thereon.

\balance
\bibliographystyle{plain}

\bibliography{local,bibliography}

\end{document}

%% file: preamble.tex
\newcommand{\tuple}[3][\ ]{{\tt #2}{#1}({#3})}

\newcommand{\clos}[1]{\tuple{clos}{#1}}

\newcommand{\evalone}{{\tt ev}}
\newcommand{\apalone}{{\tt ap}}
\newcommand{\ev}[2][\ ]{\tuple[#1]{ev}{#2}}
\newcommand{\co}[2][\ ]{\tuple[#1]{co}{#2}}
\newcommand{\ap}[2][\ ]{\tuple[#1]{ap}{#2}}

\newcommand{\kfnalone}{{\tt fun}}
\newcommand{\kmt}{{\tt halt}}
\newcommand{\kar}[2][\ ]{\tuple[#1]{arg}{#2}}
\newcommand{\kfn}[2][\ ]{\tuple[#1]{fun}{#2}}
\newcommand{\kif}[2][\ ]{\tuple[#1]{ifk}{#2}}

\newcommand{\syntax}[1]{{\tt #1}}
\newcommand{\sapp}[3][\ ]{\tuple[#1]{app}{#2,#3}}
\newcommand{\slam}[3][\ ]{\tuple[#1]{lam}{#2,#3}}

\newcommand{\svar}[2][\ ]{\tuple[#1]{var}{#2}}

\newcommand{\saddr}[1]{\tuple{addr}{#1}}
\newcommand{\superposition}[1]{\tuple{sp}{#1}}

\newcommand{\spchoice}[2]{#1}

\newcommand{\sif}[4][\ ]{\tuple[#1]{if}{#2,#3,#4}}

\newcommand{\strue}{{\tt tt}}
\newcommand{\sfalse}{{\tt ff}}
\newcommand{\saddone}{{\tt add1}}
\newcommand{\ssubone}{{\tt sub1}}
\newcommand{\szerohuh}{\syntax{zero?}}
\newcommand{\szero}{\syntax{0}}
\newcommand{\slit}[2][\ ]{\tuple[#1]{lit}{#2}}

\newcommand{\sNum}{\syntax{Z}}

\newcommand{\ext}[3]{#1\sqcup[#2\mapsto#3]}

\newcommand{\maddr}{a}
\newcommand{\mkaddr}{a_\kappa}

\newcommand{\mvar}{x}

\newcommand{\mexp}{e}
\newcommand{\mexpi}[1]{e_{#1}}

\newcommand{\menv}{\rho}
\newcommand{\mkont}{\kappa}
\newcommand{\msto}{\sigma}
\newcommand{\mop}{o}
\newcommand{\mval}{v}
\newcommand{\mvalstack}{V}
\newcommand{\mnum}{z}
\newcommand{\mbln}{b}
\newcommand{\mvalx}[1]{#1}
\newcommand{\machstep}{\mathbin{\longmapsto}}
\newcommand{\multimachstep}{\longmapsto\!\!\!\!\!\rightarrow}
\newcommand{\mlit}{l}
\newcommand{\mstate}{\varsigma}
\newcommand{\mcomp}{k}
\newcommand{\mcompi}[1]{\mcomp_{#1}}
\newcommand{\interpdelta}{\Delta}
\newcommand{\msdiff}{\xi}

\newcommand{\mseen}{S}
\newcommand{\mseentime}{\hat{S}}

\newcommand{\compile}[1]{\llbracket#1\rrbracket}
\newcommand{\kpush}[2][\ ]{\lfloor #2 \rfloor{#1}}

\newcommand{\mlab}{{\ell}}
\newcommand{\mcntr}{{t}}
\newcommand{\mtcntr}{{t_0}}

\newcommand{\mtlst}{\epsilon}
\newcommand{\cons}[2]{#1 {\tt :} #2}
\newcommand{\ttuple}[2]{(#1, #2)} 
\newcommand{\Set}{\mathcal{P}}

\newcommand{\eval}{\mathit{eval}}
\newcommand{\traces}{\mathit{traces}}
\newcommand{\alloc}{\mathit{alloc}}
\newcommand{\allockont}{\mathit{allockont}}
\newcommand{\tick}{\mathit{tick}}
\newcommand{\replay}{\mathit{replay}}
\newcommand{\diffp}{\mathit{\delta?}}

\newcommand{\append}{\mathit{append}}
\newcommand{\appendall}{\mathit{appendall}}

\newcommand{\wn}{\mathit{wn}}

\newcommand{\force}{\mathit{force}}

\newcommand{\inject}{\mathit{inject}}
\newcommand{\lookup}{\mathit{lookup}}

\newcommand{\setof}[1]{\{#1\}}

\newcommand{\mastate}{\hat{\mstate}}

\newcommand{\Context}{\mathit{Context}}
\newcommand{\cmachstep}{\mathbin{\compile{\machstep}}}
\newcommand{\camachstep}{\mathbin{\compile{\widehat{\machstep}}}}
\newcommand{\lmachstep}{\mathbin{\longmapsto_{{\mathcal L}}}}
\newcommand{\dmachstep}{\mathbin{\longmapsto_{\mathit{\msto\msdiff}}}}

\newcommand{\damachstep}{\mathbin{\widehat{\longmapsto}_{\mathit{\msto\msdiff}}}}

\newcommand{\hd}{\mathit{hd}}

\newcommand{\Label}{\mathit{Label}}
\newcommand{\Counter}{\mathit{Time}}
\newcommand{\System}{\mathit{System}}

\newcommand{\erankt}{\mathit{erankt}}
\newcommand{\erankl}{\mathit{erankl}}

\newcommand{\Store}{\mathit{Store}}
\newcommand{\StoreDelta}{\mathit{Store\Delta}}
\newcommand{\Addr}{\mathit{Addr}}
\newcommand{\Value}{\mathit{Value}}
\newcommand{\Var}{\mathit{Var}}
\newcommand{\Env}{\mathit{Env}}
\newcommand{\Expr}{\mathit{Expr}}
\newcommand{\Kont}{\mathit{Kont}}
\newcommand{\State}{\mathit{State}}
\newcommand{\Storeable}{\mathit{Storeable}}
\newcommand{\Timestamp}{{\mathbb N}} 
\newcommand{\Valstack}{\mathit{ValStack}}

\newcommand{\Boolean}{\mathit{Boolean}}

\newcommand{\changep}{\Delta\mbox{{\tt ?}}}

\newcommand{\Church}{Vardoulakis and Shivers}

%% file: bench-overview.tex
\begin{tabular}{@{}l||r||r|r||r|r||r|r@{}}
Program & LOC
& \multicolumn{2}{c||}{Time {\small (sec)}}
& \multicolumn{2}{c||}{Space {\small (MB)}}
& \multicolumn{2}{c@{}}{Speed {\small $\frac{state}{sec}$}}
\\
\hline\hline
nucleic & 3492 & \text{{\small $m$}} & 66.9 & \text{{\small $m$}} & 238 & 44 & 9K \\
matrix & 747 & \text{{\small $t$}} & 3.4 & 294 & 114 & 68 & 87K \\
nbody & 1435 & \text{{\small $t$}} & 22.9 & 361 & 171 & 67 & 57K \\
earley & 667 & 1.1K & 0.4 & 409 & 114 & 252 & 95K \\
maze & 681 & \text{{\small $t$}} & 2.6 & 332 & 114 & 55 & 118K \\
church & 42 & 44.9 & 0.1 & 86 & 114 & 714 & 56K \\
lattice & 214 & 348.5 & 0.2 & 231 & 114 & 382 & 104K \\
boyer & 642 & \text{{\small $m$}} & 13.4 & \text{{\small $m$}} & 130 & 39 & 39K \\
mbrotZ & 69 & 373.6 & 0.1 & 295 & 114 & 540 & 63K
\end{tabular}